\documentclass{article}

\usepackage{amssymb}

\usepackage{latexsym,amsmath,amsfonts,amscd}
\usepackage{xfrac}
\usepackage{graphics}
\usepackage{caption}
\usepackage{subcaption}
\usepackage{epstopdf}
\usepackage{color}
\usepackage{changebar}
\usepackage{indentfirst}
\usepackage{verbatim}
\usepackage{xcolor}
\usepackage{hyperref}
\usepackage{amssymb}
\usepackage{lscape}
\usepackage{array,multirow}
\usepackage[titletoc,title]{appendix}

\renewcommand{\d}{{\rm d}}			
\newcommand{\CFL}{\mathsf{CFL}}		

\makeatletter
\newcounter{savesection}
\newcounter{apdxsection}
\renewcommand\appendix{\par
  \setcounter{savesection}{\value{section}}%
  \setcounter{section}{\value{apdxsection}}%
  \setcounter{subsection}{0}%
  \gdef\thesection{Appendix \@Alph\c@section}}
\newcommand\unappendix{\par
  \setcounter{apdxsection}{\value{section}}%
  \setcounter{section}{\value{savesection}}%
  \setcounter{subsection}{0}%
  \gdef\thesection{\@arabic\c@section}}
\makeatother

\usepackage{authblk}
\usepackage[margin=3.0cm]{geometry}

\usepackage{lineno}

\title{Computational hemodynamics in arteries with the one-dimensional augmented fluid-structure interaction system: viscoelastic parameters estimation and \\comparison with in-vivo data}

\author[$\dagger$]{Giulia Bertaglia \footnote{Corresponding author. Email address: \textit{giulia.bertaglia@unife.it}}}
\author[$\star$]{Adri\'an Navas-Montilla}
\author[$\dagger$]{Alessandro Valiani}
\author[$\ddag$]{Manuel Ignacio Monge Garc\'ia}
\author[$\star$]{Javier Murillo}
\author[$\dagger$]{Valerio Caleffi}

\affil[$\dagger$]{\small Department of Engineering, University of Ferrara, Via G. Saragat 1, 44122 Ferrara, Italy}
\affil[$\star$]{\small Fluid Mechanics, LIFTEC, CSIC-Universidad de Zaragoza, Calle Mar\'ia de Luna 3, 50018 Zaragoza, Spain}
\affil[$\ddag$]{\small Intensive Care Unit, Hospital SAS de Jerez, C/ Circunvalaci\'on, 11407 Jerez de la Frontera, Spain}

\begin{document}
\maketitle

\begin{abstract}
Mathematical models are widely recognized as a valuable tool for cardiovascular diagnosis and the study of circulatory diseases, especially to obtain data that require otherwise invasive measurements. To correctly simulate body hemodynamics, the viscoelastic properties of vessels walls are a key aspect to be taken into account as they play an essential role in cardiovascular behavior. The present work aims to apply the augmented fluid-structure interaction system of blood flow to real case studies to assess the validity of the model as a valuable resource to improve cardiovascular diagnostics and the treatment of pathologies. First, the ability of the model to correctly simulate pulse waveforms in single arterial segments is verified using literature benchmark test cases. Such cases are designed taking into account a simple elastic behavior of the wall in the upper thoracic aorta and in the common carotid artery. Furthermore, in-vivo pressure waveforms, extracted from tonometric measurements performed on four human common carotid arteries and two common femoral arteries, are compared to numerical solutions. It is highlighted that the viscoelastic damping effect of arterial walls is required to avoid an overestimation of pressure peaks. An effective procedure to estimate the viscoelastic parameters of the model is herein proposed, which returns hysteresis curves of the common carotid arteries dissipating energy fractions in line with values calculated from literature hysteresis loops in the same vessel.
\end{abstract}


\section{Introduction}
The potential of mathematical models to support research in the field of hemodynamics and cardiovascular medicine has been widely recognized \cite{formaggia2009,alastruey2007,liang2018,muller2014,muller2019}. In recent years, blood flow models have been continuously developed, focusing on several key issues, significant to adequately represent the human circulation network. The cyclic ejection of blood from the heart results in pressure and flow pulses, transmitted along vascular pathways, with varying shape, amplitude and mean values. These variations are determined by the physical and mechanical properties of blood and vessels walls, which are the essence of a complex fluid-structure interaction (FSI) mechanism, as well as by the anatomy of the entire cardiac network \cite{nichols2011,wang2016}. Viscoelastic properties of vessels play an essential role in the cardiovascular behavior \cite{salvi2012,nichols2011,holenstein1980}. In fact, viscoelasticity is one of the features that must be realistically included in the mathematical model when accurate numerical results are sought \cite{holenstein1980,alastruey2011,montecinos2014}. Vessel walls manifest viscoelastic properties that are summed up in three main attributes: creep, stress relaxation and hysteresis \cite{battista2015,salvi2012,lakes2009}. Among the existing linear viscoelastic models, the Standard Linear Solid (SLS) model provides a better representation of the arterial wall mechanics than the generally adopted Kelvin-Voigt model \cite{alastruey2011,montecinos2014,wang2014,
mynard2015,liang2018}, being the latter unable to describe an exponential decay of stress over time \cite{westerhof2004,bessems2008,valdez-jasso2009}. On the other hand, when modeling the vessel mechanics by means of an elastic behaviour, the information related to hysteresis (i.e. the energy dissipated by viscoelastic effects) vanishes
and pressure peaks are overestimated \cite{alastruey2011,battista2015,holenstein1980}.\par
The augmented FSI (a-FSI) system for blood flow modeling, presented in \cite{bertaglia2019a,bertaglia2019b}, is herein extended to real case studies in single arteries, to assess the capability of the model to serve as a valuable tool for practical medical applications, cardiovascular diagnosis and the study of circulatory pathologies. The extension of the model underlines the importance of accounting for the viscoelastic characterization of the arterial walls. A preliminary effective strategy to estimate the viscoelastic parameters characterizing the SLS model is proposed. With this aim, in-vivo flow velocity and pressure measurements of human common carotid arteries (CCA) and common femoral arteries (CFA) are performed to assess the validity of the model, together with literature hysteresis curves of different CCAs \cite{giannattasio2008,salvi2012}.\par
The paper is structured as follows. Section \ref{section_methods} presents the one-dimensional a-FSI system of equations for blood flow, focusing on viscoelastic characterization and friction losses, and the 3-element Windkessel model chosen to reproduce peripheral wave propagation. The Implicit-Explicit (IMEX) Runge-Kutta (RK) scheme adopted to solve the system is briefly discussed and inlet and outlet boundary conditions are clearly defined. Furthermore, the methodology applied for Doppler and tonometer data acquisition and extrapolation is discussed and the procedure to estimate the viscoelastic parameters of the SLS model coupled with the governing equations of the system is presented. Numerical results are reported and discussed in Section \ref{section_results}, starting from literature benchmark tests for the solely elastic rheological characterization of the vessel wall in the upper thoracic aorta (TA) and in the CCA, followed by CFA and CCA tests designed recurring to in-vivo measurements, to validate viscoelastic solutions. Finally, Section \ref{section_conclusions} consists of a summary and concluding remarks.
\section{Methods}
\label{section_methods}
The model here presented is an extension of the blood flow model described in \cite{bertaglia2019a,bertaglia2019b}, to which the reader is referred for further details.
\subsection{1D mathematical model: the a-FSI system}
\label{section1D}
The one-dimensional governing equations for blood flow in medium to large-size vessels are obtained by integrating the 3D Navier-Stokes equations over the cross-section of the vessel, assuming axial symmetry of the system \cite{formaggia2009}. To close the resulting system, formed by the equations of conservation of mass and momentum, a tube law is needed, whereby the internal pressure is related to the cross-sectional area. Generally, a simple elastic relationship is chosen, obtaining a first fair approximation of the real behavior of vessels \cite{boileau2015,liang2009a,matthys2007,muller2014a,mynard2008,sherwin2003,xiao2014,willemet2014}. To better characterize the FSI occurring between blood and vessel wall, a viscoelastic constitutive model is required \cite{alastruey2011,alastruey2012a,nichols2011,reymond2009,valdez-jasso2009}. If the tube law is included in the system of equations in the form of a partial differential equation (PDE), the so-called a-FSI system for blood flow is obtained \cite{bertaglia2019a,bertaglia2019b}:
\begin{subequations}
\begin{align}
	&\partial_t A + \partial_x (Au) = 0 \label{eq.cont}\\
	&\partial_t (Au) + \partial_x (Au^2)  + \frac{A}{\rho} \hspace{0.5mm} \partial_x p = \frac{f}{\rho} 		 \label{eq.mom}\\
	&\partial_t p + d \hspace{1mm}\partial_x (Au) = S \label{eq.PDE}\\
	&\partial_t A_0 = 0 \label{eq.A0} \\
	&\partial_t E_0 = 0 \label{eq.E0} \\
	&\partial_t p_{ext} = 0 \label{eq.pext} \hspace{0.5mm}.
\end{align}
\label{completesyst}
\end{subequations}
Here $A(x,t)$ is the cross-sectional area of the vessel, $u(x,t)$ is the cross-section averaged blood velocity, $p(x,t)$ is the internal blood pressure, $\rho$ is the blood density, $f$ is the friction loss term and $x$ and $t$ are space and time respectively. To cope with possible longitudinal discontinuities of geometrical and mechanical properties of vessels, such as equilibrium cross-sectional area $A_0(x)$, instantaneous Young modulus $E_0(x)$ and external pressure $p_{ext}(x)$, additional equations~\eqref{eq.A0},~\eqref{eq.E0},~\eqref{eq.pext} are introduced in the system to allow a formally correct numerical treatment \cite{muller2013}. In the constitutive PDE~\eqref{eq.PDE}, the parameter $d(A,A_0,E_0)$ represents the elastic contribution of the vessel wall, having the same formulation if describing the mechanics of the wall with an elastic or with the SLS viscoelastic model \cite{lakes2009,bertaglia2018}:
\begin{equation}
d = \frac{K}{A} \left(m \alpha^m - n \alpha^n\right) ,
\label{d}
\end{equation}
where $\alpha = \sfrac{A}{A_0}$ is the non-dimensional cross-sectional area, $K(x)$ represents the stiffness coefficient of the material and $m$ and $n$ are parameters associated to the specific behavior of the vessel wall, whether arterial or venous. When dealing with arteries (as in the present study), the characterization of these parameters leads to:
\begin{equation}
K = \frac{E_0 h_0}{R_0}, \quad m = 1/2, \quad n = 0 \hspace{0.5mm},
\label{K}
\end{equation}
with $h_0$ wall thickness and $R_0$ equilibrium inner radius of the vessel.
For further details, also concerning veins, see \cite{muller2013,shapiro1977}. The source term $S(x,t)$ in Eq.~\eqref{eq.PDE} carries the viscous information of the vessel wall behavior; hence, $S = 0$ if the elastic behavior is considered, or 
\begin{equation}
S = \frac{1}{\tau_r}\left[\frac{E_{\infty}}{E_0} K \left(\alpha^m - \alpha^n\right) - (p-p_{ext})\right] 
\label{source}
\end{equation}
if the viscoelastic behavior is chosen. In the SLS model $E_0$ is the Young modulus of the first spring in series with the Kelvin-Voigt unit (formed by a single spring and a dashpot connected in parallel), $E_2$ is the Young modulus of the elastic spring of the Kelvin-Voigt element itself and $\eta$ is the viscosity coefficient of the dashpot. It follows that the asymptotic Young modulus, $E_{\infty}$, and the relaxation time, $\tau_r$, are \cite{lakes2009}:
\begin{equation}
E_\infty = \frac{E_0 E_2}{E_0 + E_2}, \qquad \tau_r = \frac{\eta}{E_0 + E_2} \hspace{0.5mm}.
\label{Einf&tau}
\end{equation}\par
Regarding friction losses, the velocity profile is considered self-similar and axisymmetric. The typical velocity profile used for blood flow satisfying the no-slip condition is \cite{alastruey2012}:
\begin{equation}
v(x,r,t) = u \frac{\zeta+2}{\zeta}\left[1-\left(\frac{r}{R}\right)^{\zeta}\right] ,
\label{velprof}
\end{equation}
where $r$ is the radial coordinate, $R(x,t)$ is the inner radius of the vessel and $\zeta = \frac{2-\alpha_c}{\alpha_c-1}$ is a parameter depending on $\alpha_c$, the Coriolis coefficient. Eq.~\eqref{velprof} defines different profiles between close to flat ($\alpha_c \approx 1$) to parabolic ($\alpha_c = \sfrac{3}{4}$, $\zeta = 2$). The velocity profile is on average rather blunt in central arteries \cite{quarteroni2004}, with the consequence that the choice of $\alpha_c = 1.1$ ($\zeta = 9$) provides the best compromise to fit experimental data \cite{xiao2014}. A parabolic velocity profile is more suitable for non-central arteries. For the velocity profile given by Eq.~\eqref{velprof}, the friction term finally results:
\begin{equation}
f = -2(\zeta+2)\mu\pi u \hspace{0.5mm},
\label{frictionterm}
\end{equation}
where $\mu$ is the dynamic viscosity of blood.
\begin{figure}[t]
\centering
\includegraphics[width=0.95\linewidth]{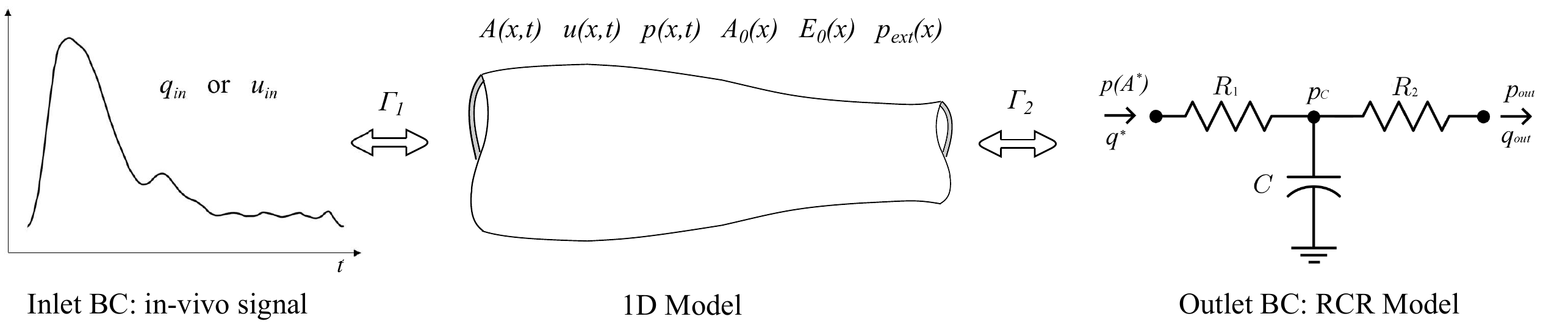}
\caption{Schematic representation of the model. Inlet boundary condition given by $q_{in}$ or $u_{in}$ and recurring to the first Riemann invariant $\Gamma_1$, defined in Eq.~\eqref{BC_RI1}. The 1D a-FSI model is coupled to the 3-element Windkessel model (analogous to the RCR electric circuit) at the outlet through the solution of a Riemann problem at the 1D/0D interface, recurring to the second Riemann invariant $\Gamma_2$, defined in Eq.~\eqref{BC_RI2}.}
\label{fig.model}
\end{figure}
\subsection{0D mathematical model: the 3-element Windkessel model}
\label{section0D}
The 0D model used to simulate the effects of peripheral resistance and compliance on pulse waves propagation in large 1D arteries is the so-called RCR or 3-element Windkessel model, which consists of a resistor, with resistance $R_1$, connected in series with a parallel combination of a second resistor, with resistance $R_2$, and a capacitor, with compliance $C$, as shown in Fig.~\ref{fig.model} \cite{boileau2015,alastruey2012,reymond2009,willemet2016,xiao2014}. In this model the inductance is neglected since peripheral inertias have a minor effect on reflected waves under normal conditions \cite{alastruey2008}. Referring to Fig.~\ref{fig.model} for the nomenclature, the final system of equations for the 0D model reads:
\begin{subequations}
\begin{align}
	&C \hspace{0.5mm} \d_t p_{C} = q^* - q_{out} \label{0D_1} \\ 
	&R_1 \hspace{0.5mm} q^* = p(A^*) - p_{C} \label{0D_2} \\
	&R_2 \hspace{0.5mm} q_{out} = p_{C} - p_{out} \hspace{0.5mm}, \label{0D_3}
\end{align}
\label{RCRsyst}
\end{subequations}
where $p_C$ is the pressure at the capacitor and $q^* = A^*u^*$. Resistance $R_1$ is introduced to absorb incoming waves and reduce artificial backward reflections in large arteries, so it is fixed to match the characteristic impedance $Z_0$ of the terminal 1D vessel \cite{alastruey2008}: $Z_0 = \sfrac{\rho c_0}{A_0}$.
\subsection{Numerical model}
\label{section_numericalmodel}
To solve system~\eqref{completesyst}, the IMEX-RK scheme proposed in \cite{pareschi2005} for applications to hyperbolic systems with stiff relaxation terms is considered. The stiffness of system~\eqref{completesyst} depends on the relaxation time of the material, $\tau_r$, of the source term in Eq.~\eqref{source}. The adopted scheme is asymptotically preserving and asymptotically accurate in the zero relaxation limit. Following \cite{bertaglia2019a,bertaglia2019b}, a formally implicit finite volume discretization is applied, using a second-order L-stable diagonally implicit RK method for the stiff part, which gives elevated robustness to the method; for the non-stiff terms, a second-order explicit strong-stability-preserving (SSP) method is employed, considering the path-conservative Dumbser-Osher-Toro (DOT) Riemann solver \cite{bertaglia2018,dumbser2011a,dumbser2011}. For the system of equations here discussed, a totally explicit algorithm is derived, leading to a consistent computational costs reduction.\\
The implementation of boundary conditions is discussed in detail in \ref{section_BC}.
\subsection{Human in-vivo blood velocity and pressure waveforms: data acquisition and extrapolation}
\label{section_invivodata}
The Doppler ultrasound technique (Xario 100, Toshiba Medical System, Shimoishagami, Japan) in combination with a 4.8/11 Hz linear transducer (Toshiba PLU-704BT) is used to record the time evolution of blood velocity \cite{gill1985,hwang2017} in four volunteers' CCA and two volunteers' CFA, from whom proper informed consent was obtained. A beam-flow angle of 60$^{\circ}$ and a sample volume defined by a window size of 1.0-1.5 mm are chosen for the experiments. Aware of the possible sources of error and non-optimal accuracy of Doppler ultrasound measurements \cite{blanco2015,gill1985,park2012}, this technique resulted the most applicable, still being largely adopted in literature.\\
In the post-processing, a threshold segmentation of the raw velocity is first applied to extract the maximum envelope of the signal. Then, considering the point of maximum derivative as starting point of a new cardiac cycle, an average of all the waves obtained for each cardiac cycle has been carried out, obtaining a representative waveform for each case study. Finally, under the assumption of a parabolic velocity profile in non-central arteries, a scaling coefficient of 0.5 is used to estimate the average cross-sectional velocity.\par
To measure pressure waveforms, the arterial applanation tonometry technique is used \cite{giannattasio2008,orourke2007,townsend2015, salvi2004,salvi2012,spronck2016}. A PulsePen tonometer (DiaTecne srl, Milan, Italy), which consists of a tonometric probe and an electrocardiography (EKG) unit, is chosen for this purpose \cite{salvi2004}.\\
In the post-processing, a detrending of the signal is applied to remove the low-frequency oscillations caused by the respiration of the subject, recurring to a low-pass filter. To carry out a phase averaging, the EKG is used as baseline signal to estimate the duration of the cardiac cycles, through a peak-detection algorithm. For all the cardiac cycles, the pressure waveform is re-sampled and averaged into a representative one. Finally, for the calibration of the phase-averaged pressure waveform, the diastolic and systolic pressure of the subject measured in the brachial artery (by means of a sphygmomanometer at the beginning of the experiment) are used as reference values: mean arterial pressure remains constant from the aorta to the peripheral arteries, as the diastolic pressure, which tends to decrease insignificantly from the center to the periphery \cite{salvi2012}. 
\subsection{FSI parameters estimation}
\label{section_calibration}
In the case of a purely elastic model, the single Young modulus $E_0$ is evaluated through a given reference celerity $c_0$, taken from \cite{muller2014a}, inverting the wave speed or Moens-Korteweg celerity equation \cite{nichols2011},
\begin{equation}
c = \sqrt{\frac{A}{\rho} \hspace{0.5mm} d } = \sqrt{\frac{E_0 h_0}{2 R_0 \rho} \sqrt{\alpha}} \hspace{0.5mm} ,
\label{cel}
\end{equation}
considering a reference state in which $A = A_0$ \cite{alastruey2012,alastruey2012a}:
\begin{equation}
\label{E_0}
E_0 = \frac{2 R_0 \rho c_0^2}{h_0} \hspace{0.5mm} .
\end{equation}
The reference state in this work is assumed coincident with the diastolic state. When the viscosity of the vessel wall is taken into account considering the SLS model, the elastic component is represented by two different Young modulus: the instantaneous, $E_0$, and the asymptotic one, $E_{\infty}$, defined in Eq.~\eqref{Einf&tau}. The elastic modulus of the system $E(x,t)$ changes in time, from the instantaneous to the asymptotic value, following the so-called relaxation function \cite{lakes2009}. How rapidly the viscoelastic material reaches the asymptotic condition depends on the third parameter of the model: the relaxation time, $\tau_r$. If viscoelastic materials are exposed to strain, indeed, stresses are accumulated and then gradually released during an exponential relaxation in time \cite{battista2015,lakes2009}. When the material undergoes cyclic loading and unloading, the energy put into the system is not totally recovered but partially dissipated, presenting an hysteresis loop. The amplitude of this loop does not only depend on the viscous parameter $\eta$, but also on the ratio $\sfrac{E_{\infty}}{E_0}$, hence on how much the behavior of the material differs from the elastic one (for which $E_{\infty} \equiv E_0$). \\
The estimation of the three viscoelastic parameters defining the SLS model aims to obtain a viscoelastic system that correctly matches with the corresponding elastic in the asymptotic case. When considering the SLS model with $\eta = 0$, the equivalent elastic system is represented by two springs in series for which the equivalent Young modulus exactly results $E_{\infty}$. Therefore, when switching from elastic to viscoelastic simulations, $E_{\infty}$ is imposed equal to the previously calculated $E_0$ in the elastic case (obtained with respect to $c_0$). For the determination of the viscosity parameter $\eta$, we referred to parameters estimated in\cite{alastruey2012a} for the viscosity coefficient $\Gamma$ of the commonly used Kelvin-Voigt model, for which the linear relationship with $\eta$ (neglecting Poisson's coefficient) reads: \cite{alastruey2011,montecinos2014,mynard2015,wang2014}:
\begin{equation}
\eta = \frac{2\Gamma}{h_0 \sqrt{\pi}} \hspace{0.5mm}.
\label{eta}
\end{equation}
Finally, the ratio $\sfrac{E_{\infty}}{E_0}$ is evaluated so that numerical results reproduce a realistic energy loss, based on published hysteresis loops obtained from in-vivo pressure-diameter measurements of human CCAs \cite{giannattasio2008,salvi2012}, shown in grey-scale in Fig.~\ref{fig.hysteresis} for subjects of different ages. The energy dissipated in literature hysteresis curves is numerically evaluated by an integral, being the area inside the loop in the $p$-$A$ plane. Observing Fig.~\ref{fig.hysteresis}, it can be noticed that there is a high variability of hysteresis loops recorded in young subjects from those in elderly subjects, this also being the case for subjects with a higher compared with a lower blood pressure level, as highlighted in \cite{giannattasio2008}. To cope with these large fluctuations of energy losses, values of energy dissipated by viscoelastic effects are converted to energy fractions dividing for the reference stored energy. The area of the triangle below the pulse pressure line, again in the $p$-$A$ plane, is assumed as reference stored energy \cite{wang2012}. The weighted average over age is then taken as reference for comparisons with simulated dissipations. The empirical relation which returned the best fit with reference data in CCAs resulted:
\begin{equation}
\frac{E_{\infty}}{E_0} = e^{-1.3 \cdot 10^{-5}\eta} \hspace{0.5mm},
\label{Eratio_calib}
\end{equation}
through which the viscoelastic $E_0$ can be imposed.
Eq.~\eqref{Eratio_calib} is used for the estimation of Young modulus ratios in all the arteries of this study due to the limited $p$-$D$ data available in literature concerning other human vessels. The proposed relation respects the elastic asymptotic limit, being $\sfrac{E_{\infty}}{E_0} = 1$ when $\eta = 0$. For a more robust estimation of viscoelastic parameters for the entire human cardiovascular network, further testing collecting additional in-vivo data from humans is appropriate. 
A schematic algorithm of the above discussed procedure is available in \ref{appendix_calib}.
\begin{figure}[t!]
\begin{subfigure}{0.5\textwidth}
\centering
\includegraphics[width=0.75\linewidth]{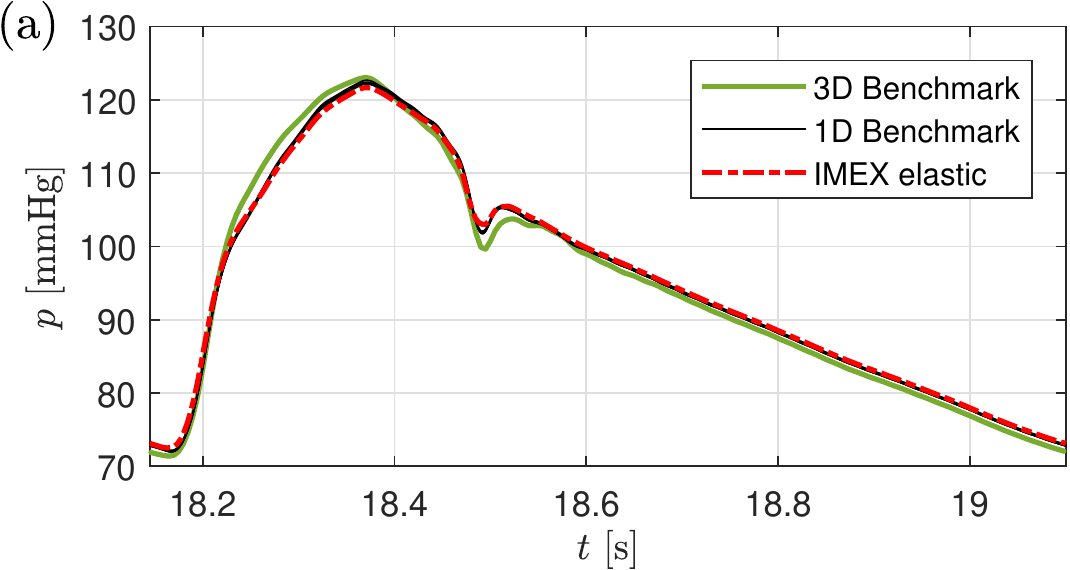}
\label{fig.BAp}
\vspace*{2mm}
\end{subfigure}
\begin{subfigure}{0.5\textwidth}
\centering
\includegraphics[width=0.75\linewidth]{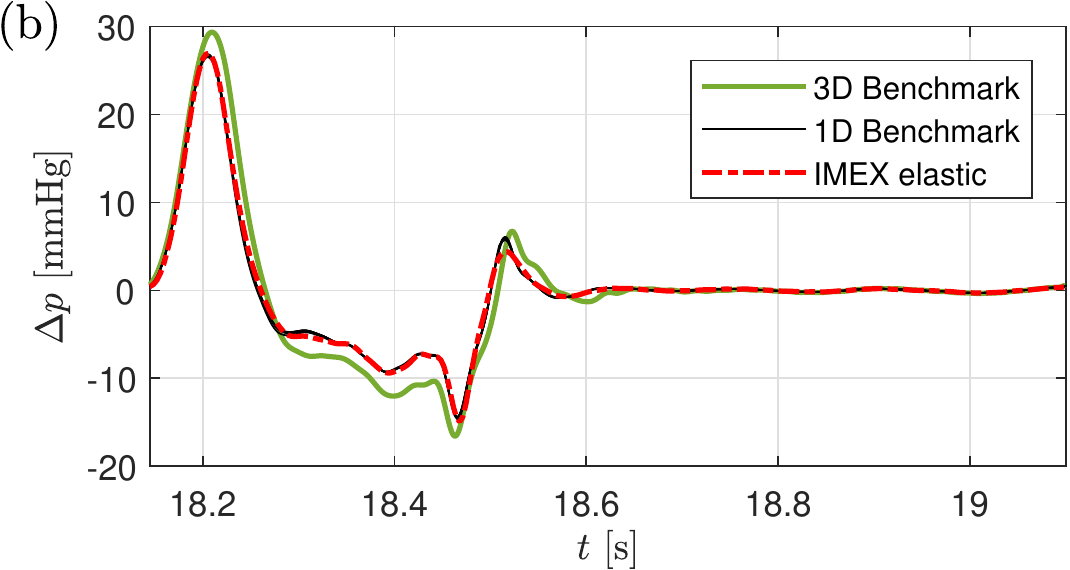}
\label{fig.BAdeltap}
\vspace*{2mm}
\end{subfigure}
\begin{subfigure}{0.5\textwidth}
\centering
\includegraphics[width=0.75\linewidth]{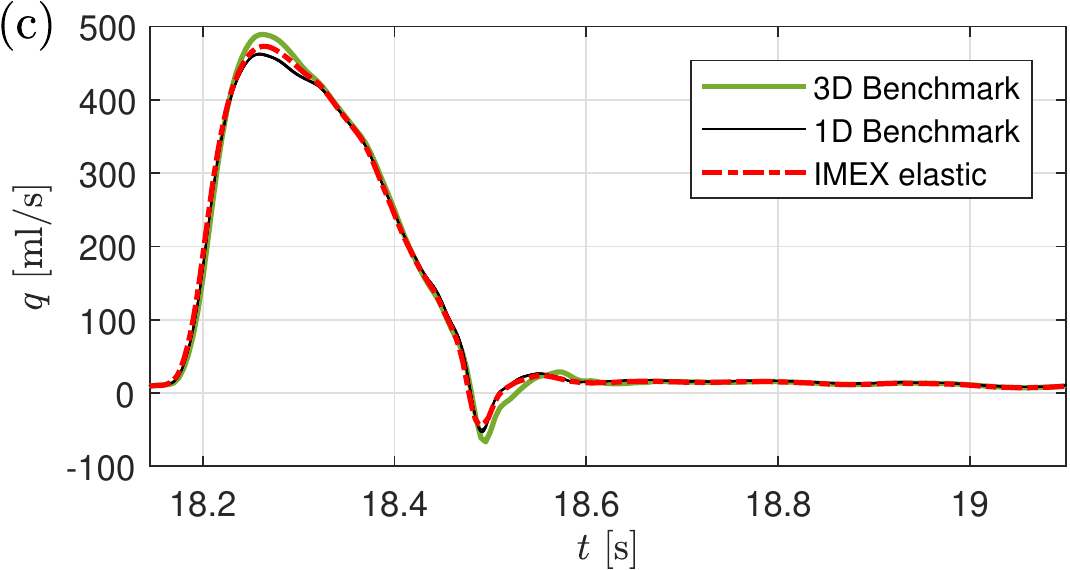}
\label{fig.BAq}
\end{subfigure}
\begin{subfigure}{0.5\textwidth}
\centering
\includegraphics[width=0.75\linewidth]{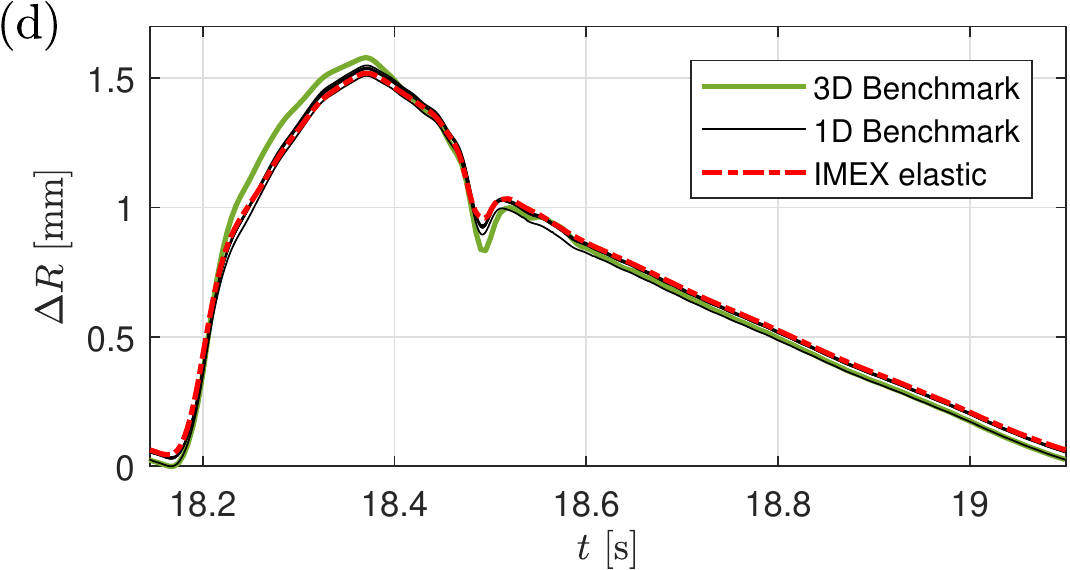}
\label{fig.BAdeltaR}
\end{subfigure}
\caption{Baseline upper thoracic aorta case (cTA). Solution of the a-FSI system with the IMEX scheme with elastic tube law compared to six 1D (discontinous Galerkin, locally conservative Galerkin, Galerkin least-squares finite elements, finite volumes, finite difference MacCormack and simplified trapezium rule method) and one 3D benchmark solutions, all taken from \cite{boileau2015}. Results presented in terms of pressure at the midpoint (a), pressure difference between inlet and outlet (b), flow rate at the midpoint (c) and change in radius from diastole at the midpoint (d).}
\label{fig.BA}
\end{figure}
\begin{figure}[t!]
\begin{subfigure}{0.33\textwidth}
\centering
\includegraphics[width=0.95\linewidth]{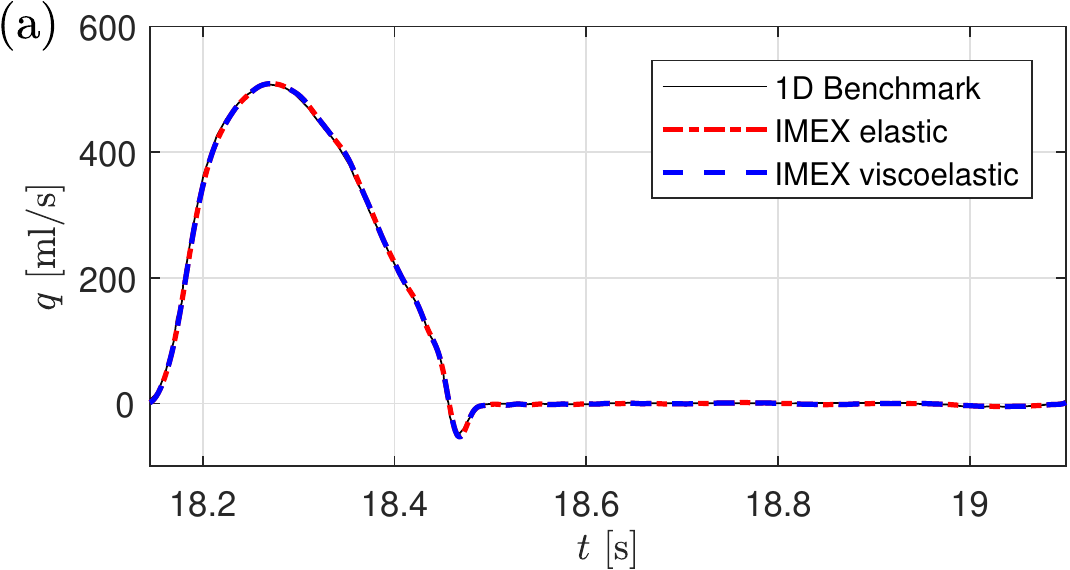}
\label{fig.TAqin}
\vspace*{2mm}
\end{subfigure}
\begin{subfigure}{0.33\textwidth}
\centering
\includegraphics[width=0.95\linewidth]{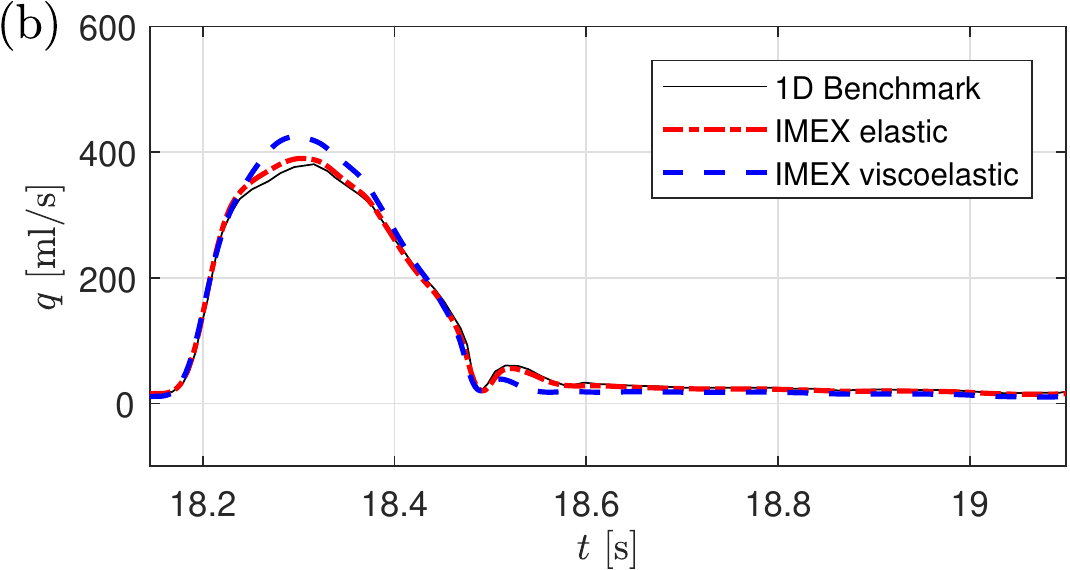}
\label{fig.TAqmid}
\vspace*{2mm}
\end{subfigure}
\begin{subfigure}{0.33\textwidth}
\centering
\includegraphics[width=0.95\linewidth]{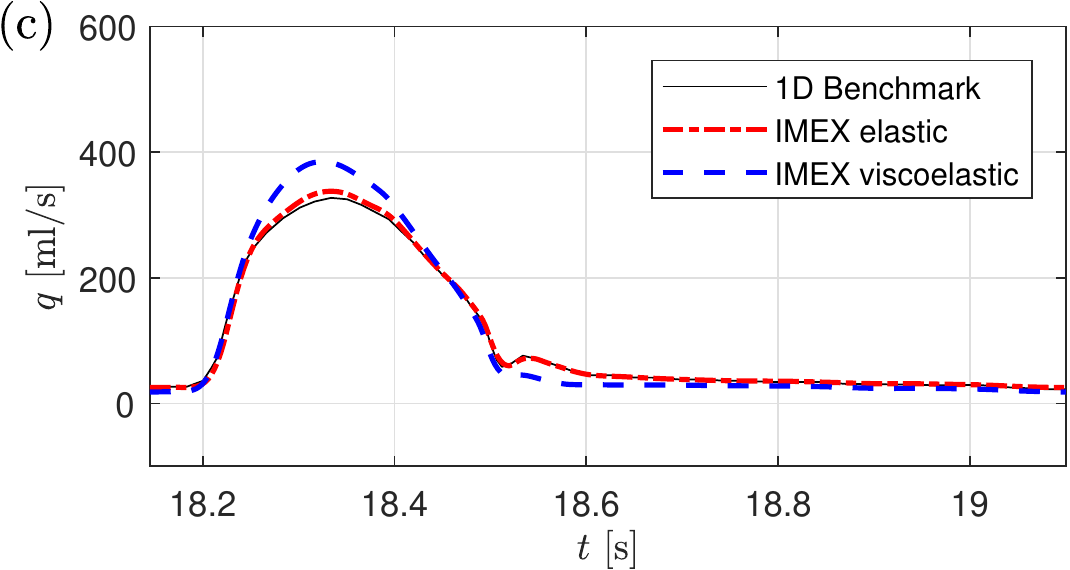}
\label{fig.TAqout}
\vspace*{2mm}
\end{subfigure}
\begin{subfigure}{0.33\textwidth}
\centering
\includegraphics[width=0.95\linewidth]{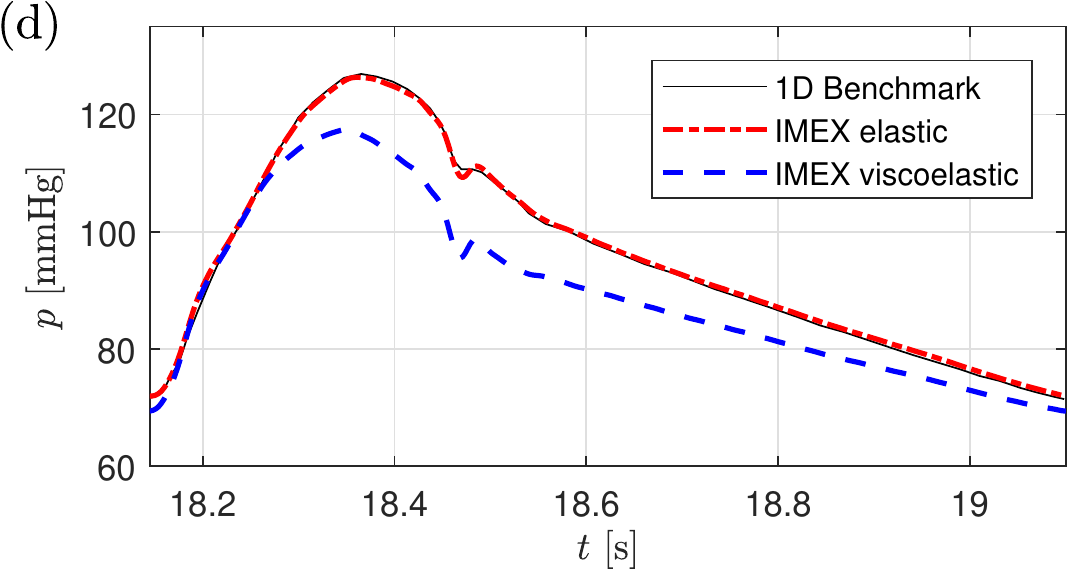}
\label{fig.TApin}
\end{subfigure}
\begin{subfigure}{0.33\textwidth}
\centering
\includegraphics[width=0.95\linewidth]{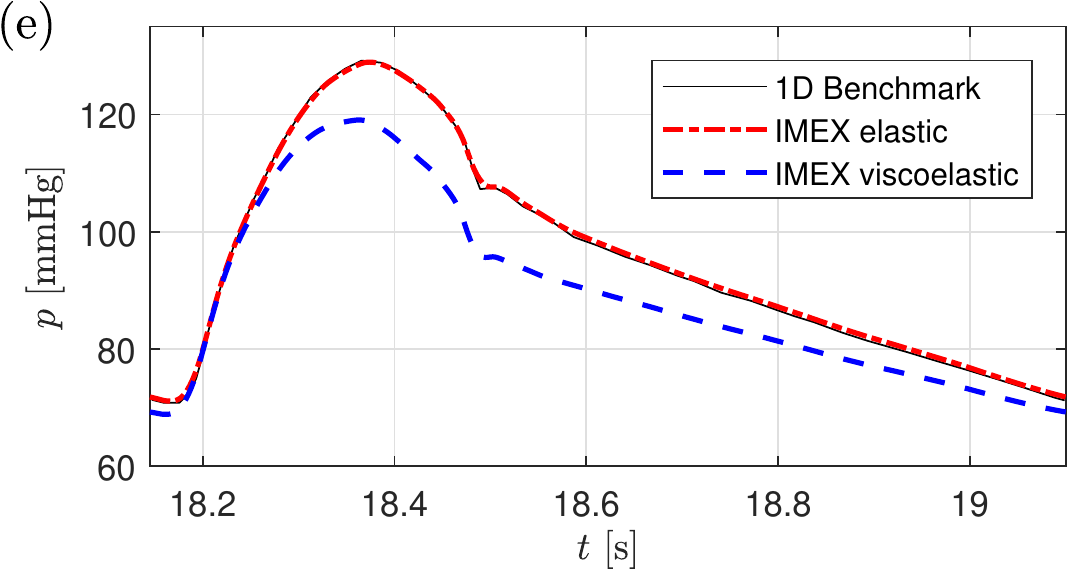}
\label{fig.TApmid}
\end{subfigure}
\begin{subfigure}{0.33\textwidth}
\centering
\includegraphics[width=0.95\linewidth]{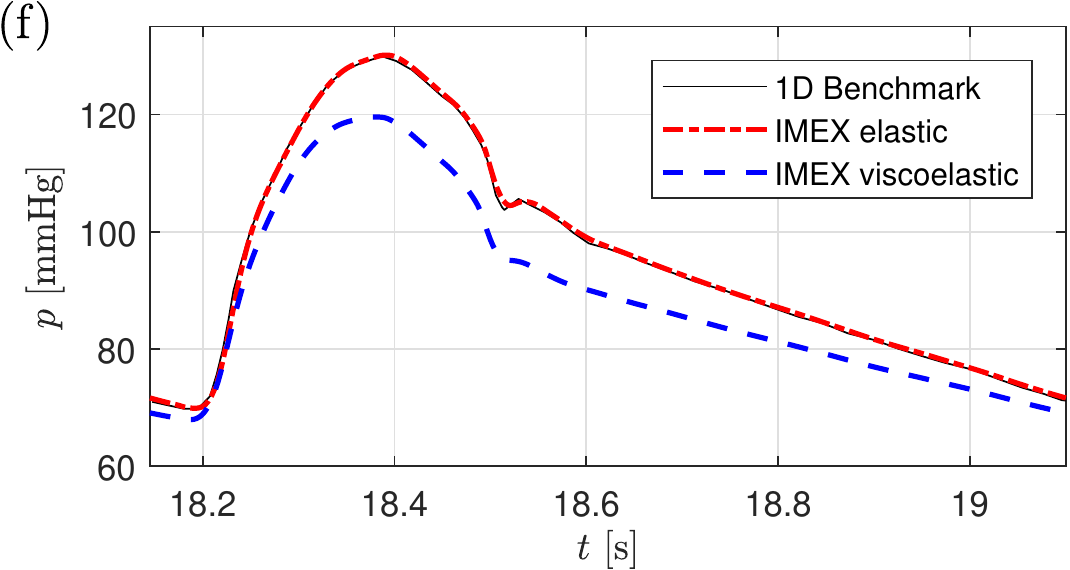}
\label{fig.TApout}
\end{subfigure}
\caption{Tapered upper thoracic aorta case (tTA). Results obtained solving the a-FSI system with the IMEX scheme with elastic and viscoelastic tube law compared to 1D elastic benchmark, taken from \cite{xiao2014}, presented in terms of flow rate at the inlet (a), flow rate at the midpoint (b), flow rate at the outlet (c), pressure at the inlet (d), pressure at the midpoint (e), pressure at the outlet (f).}
\label{fig.TA}
\end{figure}
\begin{figure}[t!]
\begin{subfigure}{0.5\textwidth}
\centering
\includegraphics[width=0.75\linewidth]{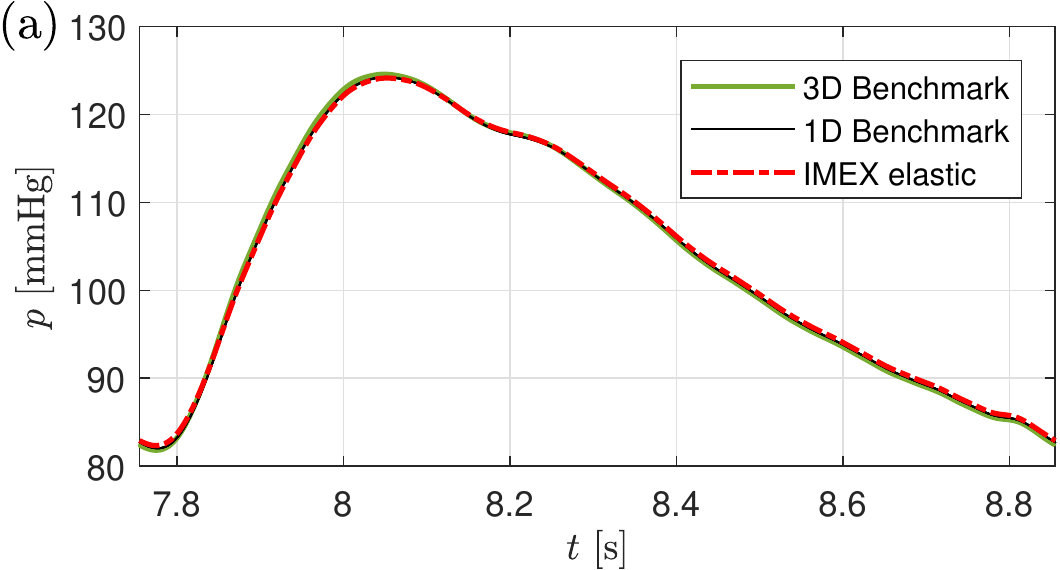}
\label{fig.BCp}
\vspace*{2mm}
\end{subfigure}
\begin{subfigure}{0.5\textwidth}
\centering
\includegraphics[width=0.75\linewidth]{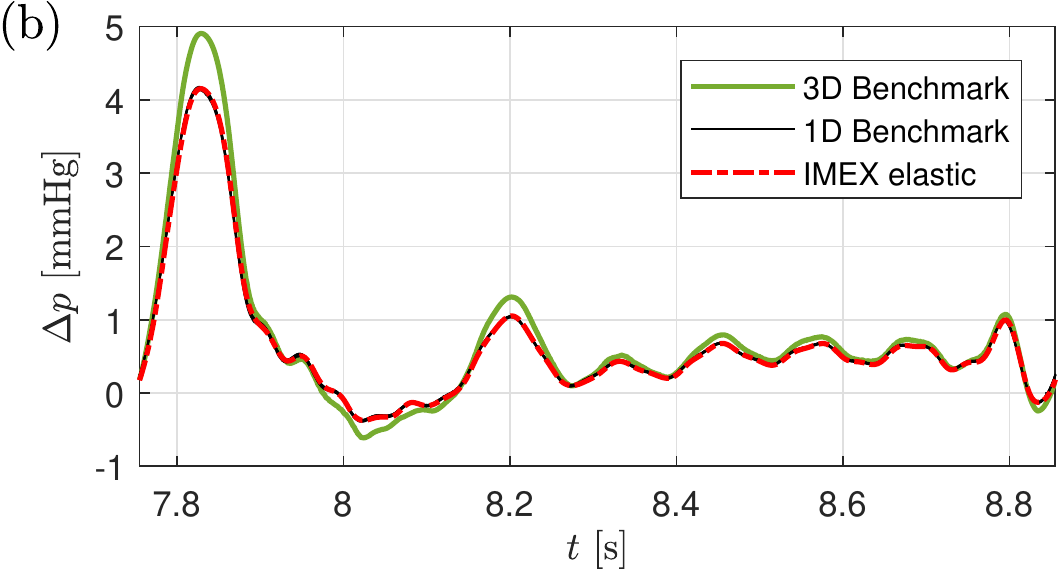}
\label{fig.BCdeltap}
\vspace*{2mm}
\end{subfigure}
\begin{subfigure}{0.5\textwidth}
\centering
\includegraphics[width=0.75\linewidth]{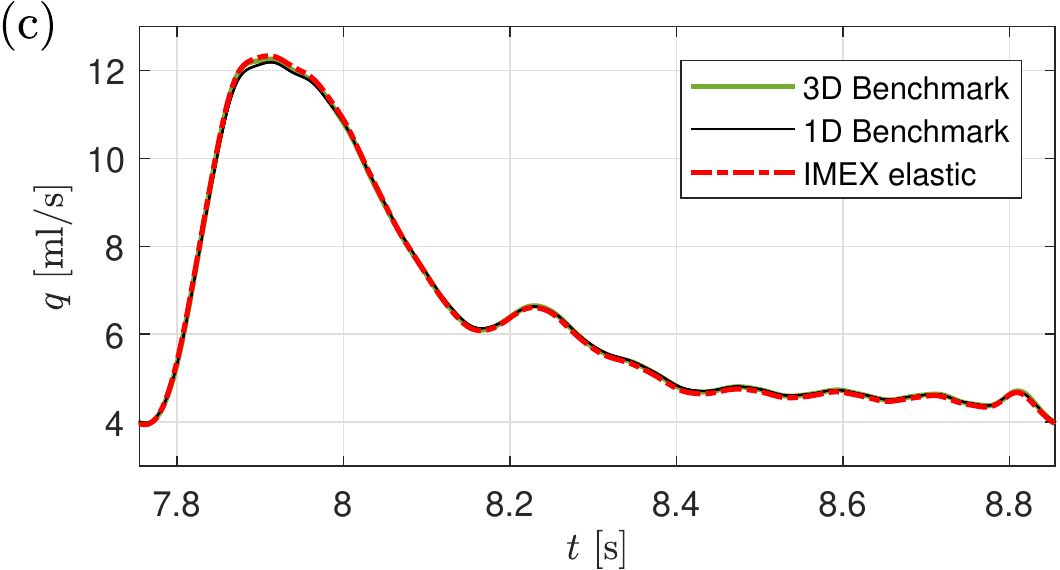}
\label{fig.BCq}
\end{subfigure}
\begin{subfigure}{0.5\textwidth}
\centering
\includegraphics[width=0.75\linewidth]{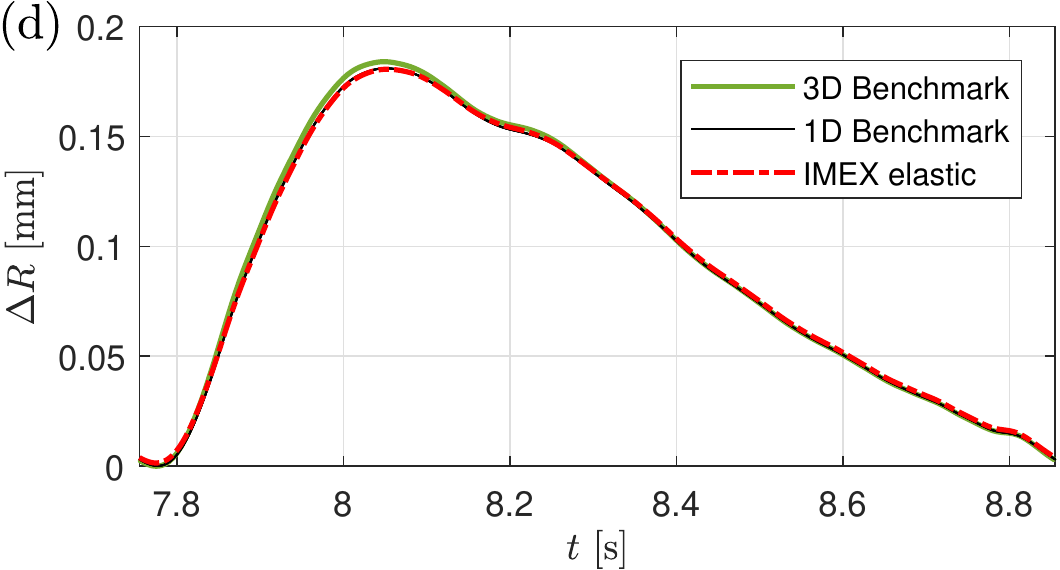}
\label{fig.BCdeltaR}
\end{subfigure}
\caption{Baseline common carotid artery case (cCCA). Solution of the a-FSI system with the IMEX scheme with elastic tube law compared to six 1D (discontinous Galerkin, locally conservative Galerkin, Galerkin least-squares finite elements, finite volumes, finite difference MacCormack and simplified trapezium rule method) and one 3D benchmark solutions, all taken from \cite{boileau2015}. Results presented in terms of pressure at the midpoint (a), pressure difference between inlet and outlet (b), flow rate at the midpoint (c) and change in radius from diastole at the midpoint (d).}
\label{fig.BC}
\end{figure}
\begin{figure}[t!]
\begin{subfigure}{0.33\textwidth}
\centering
\includegraphics[width=0.95\linewidth]{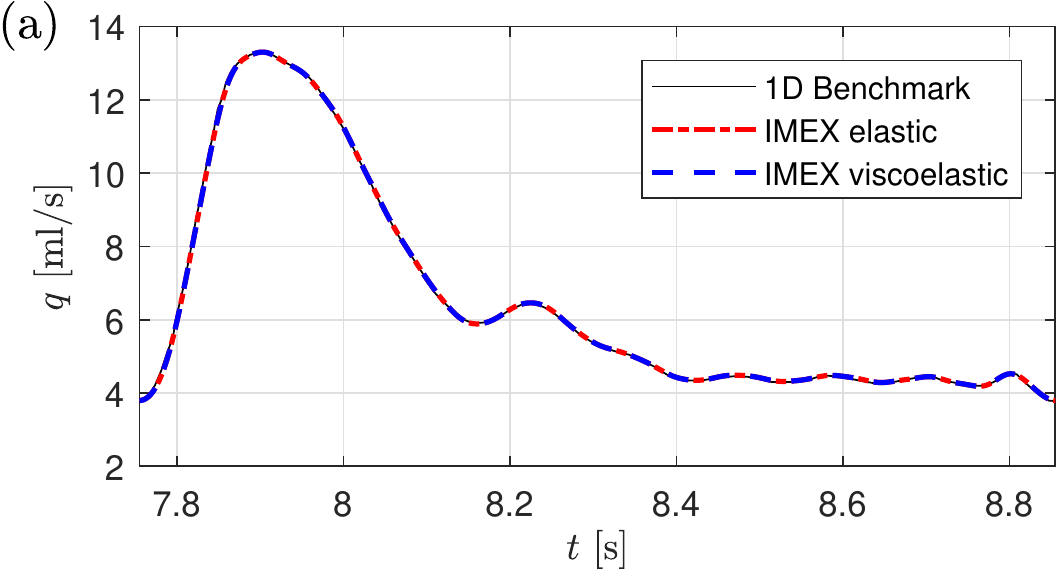}
\label{fig.TCqin}
\vspace*{2mm}
\end{subfigure}
\begin{subfigure}{0.33\textwidth}
\centering
\includegraphics[width=0.95\linewidth]{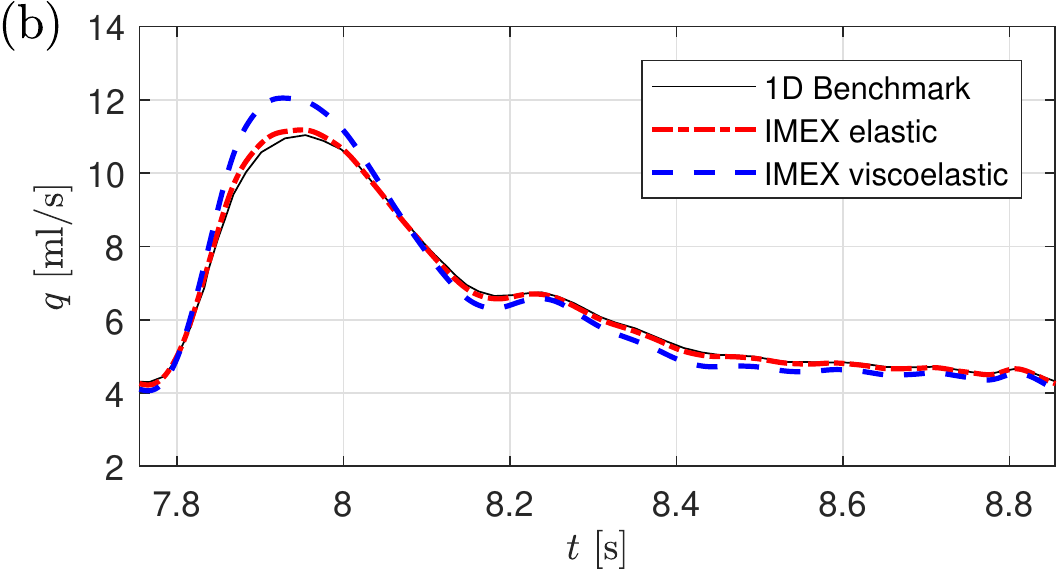}
\label{fig.TCqmid}
\vspace*{2mm}
\end{subfigure}
\begin{subfigure}{0.33\textwidth}
\centering
\includegraphics[width=0.95\linewidth]{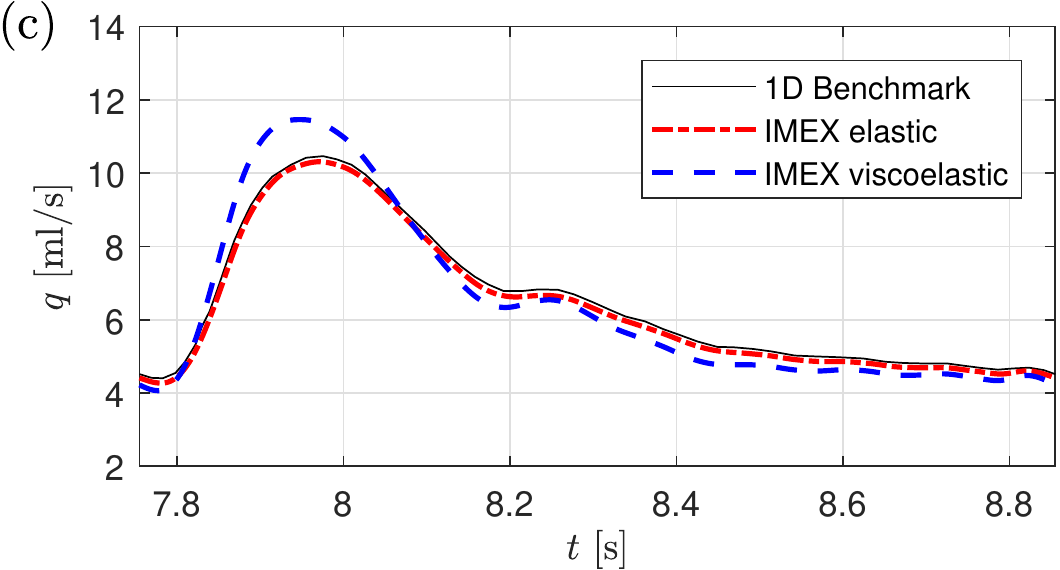}
\label{fig.TCqout}
\vspace*{2mm}
\end{subfigure}
\begin{subfigure}{0.33\textwidth}
\centering
\includegraphics[width=0.95\linewidth]{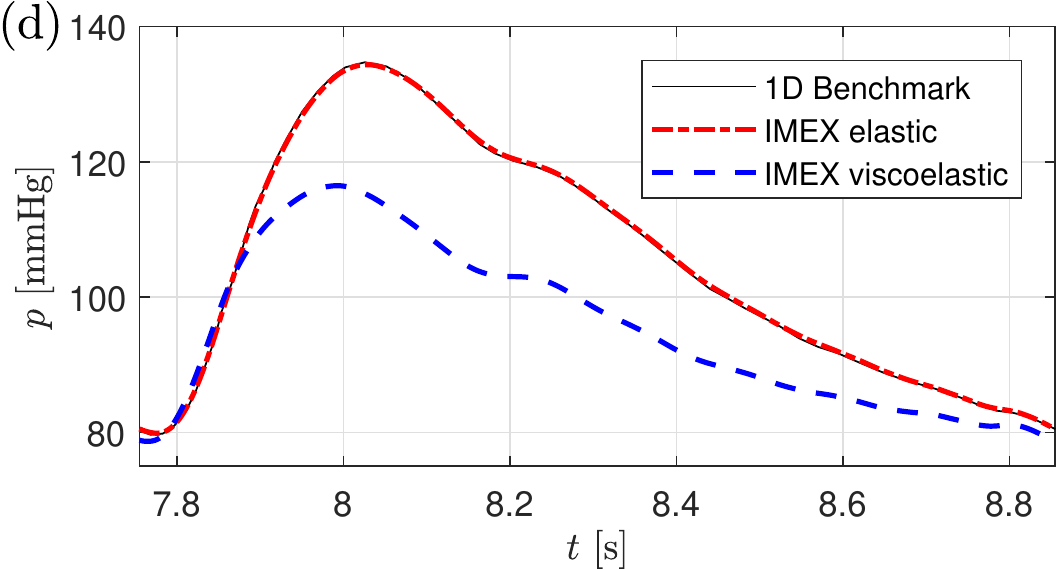}
\label{fig.TCpin}
\end{subfigure}
\begin{subfigure}{0.33\textwidth}
\centering
\includegraphics[width=0.95\linewidth]{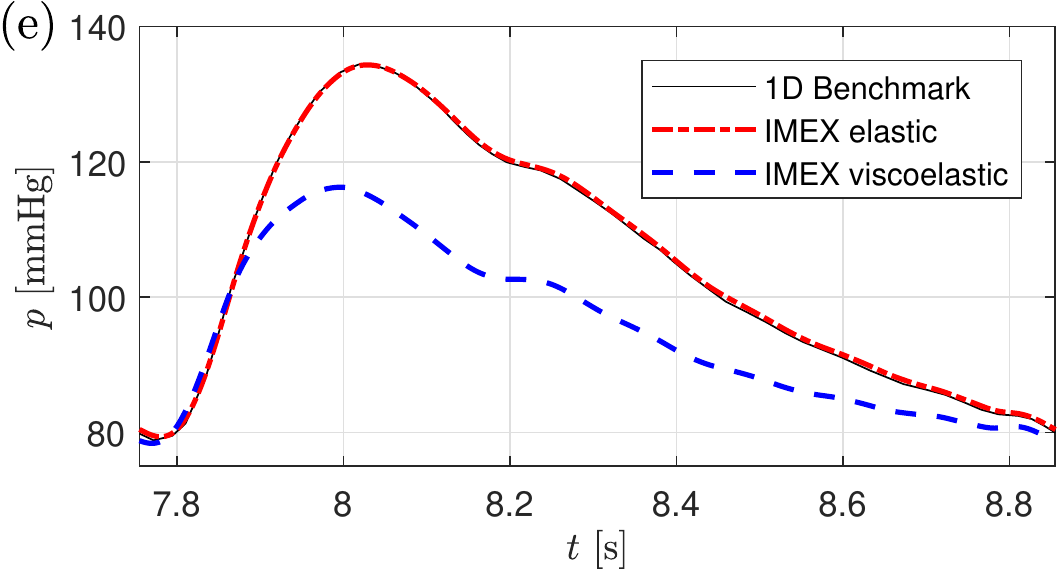}
\label{fig.TCpmid}
\end{subfigure}
\begin{subfigure}{0.33\textwidth}
\centering
\includegraphics[width=0.95\linewidth]{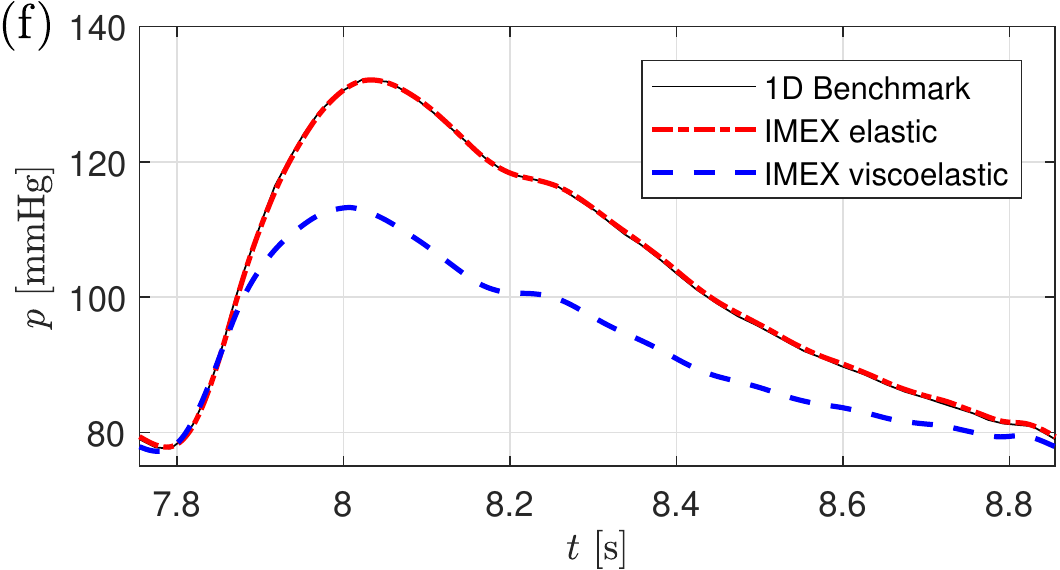}
\label{fig.TCpout}
\end{subfigure}
\caption{Tapered common carotid artery case (tCCA). Results obtained solving the a-FSI system with the IMEX scheme with elastic and viscoelastic tube law compared to 1D elastic benchmark, taken from \cite{xiao2014}, presented in terms of flow rate at the inlet (a), flow rate at the midpoint (b), flow rate at the outlet (c), pressure at the inlet (d), pressure at the midpoint (e), pressure at the outlet (f).}
\label{fig.TC}
\end{figure}
\section{Results and Discussion}
\label{section_results}
Simulations are initially performed considering the elastic tube law to compare numerical results with benchmark data sets available in literature. The upper TA and the CCA, are first analyzed considering a constant radius \cite{boileau2015} and then with a linearly tapered radius \cite{xiao2014}. Simulations for the tapered vessel cases are additionally run using the viscoelastic SLS model. To further validate the model, test cases regarding the CCA and the CFA are designed using in-vivo data discussed in Section~\ref{section_invivodata}. For these tests, simulations are performed comparing results using the elastic and the viscoelastic model, to assess the effects of wall viscoelasticity in single vessels.\\
In all the presented problems, when switching from the elastic to the viscoelastic tube law, only the viscoelastic parameters are activated, without changing other properties, including inlet and outlet boundary conditions. Outflow Windkessel parameters are calibrated following the procedure proposed in \cite{alastruey2012a,xiao2014}. Viscoelastic parameters are estimated as presented in Section~\ref{section_calibration} and \ref{appendix_calib}.\\
Model parameters for the TA and CCA benchmark test cases are given in Table~\ref{tab.dataAlastruey}, while in Table~\ref{tab.datatestinvivo} all the model parameters are listed for each in-vivo data test case.
\begin{table}[h!]
\centering
\begin{tabular}{l | c c c c }
\hline
	Parameter &cTA &tTA &cCCA &tCCA\\
	\hline
	\(L\) [cm] &24.137	&24.137	&12.60	&12.60 \\
	\(R_{0,in}\) [mm] &12.0	&15.0	&3.0	&4.0 \\
	\(R_{0,out}\) [mm] &12.0	&10.0	&3.0	&2.0 \\
	\(h_0\) [mm] &1.2	&1.2	&0.3	&0.3 \\
	\(p(x,0)\) [kPa] &0	&0	&0	&0 \\
	\(u(x,0)\) [m/s] &0	&0	&0	&0 \\
	\(\alpha_c\) [-] &1.1	&1.1	&\sfrac{4}{3}	&\sfrac{4}{3} \\
	\(p_D\) [kPa] &9.467	&9.467	&10.933	&10.933 \\
	\(p_{out}\) [kPa] &0	&0	&0	&0 \\
	\(R_{1}\) [MPa s m$^{-3}$] &11.752	&18.503	&248.75	&685.48 \\
	\(R_{2}\) [MPa s m$^{-3}$] &111.67	&104.92	&1869.7	&1433.0 \\
	\(C\) [m$^{3}$ GPa$^{-1}$] &10.163	&10.163	&0.17529	&0.17529 \\
	\(E_0\) [MPa] &0.5333	&0.7275	&0.9333	&1.7367 \\
	\(E_{\infty}\) [MPa] &-	&0.5333	&-	&0.9333 \\
	\(\eta\) [kPa s] &-	&23.884	&-	&47.768 \\
	\(\tau_r\) [s] &-	&0.009	&-	&0.013 \\
\hline
\end{tabular}
\caption{Model parameters of the upper thoracic aorta (TA) and the common carotid artery (CCA), with constant (c) or tapered (t) radius, taken from \cite{xiao2014,boileau2015}: vessel length $L$, inlet equilibrium radius $R_{0,in}$, outlet equilibrium radius $R_{0,in}$, vessel wall thickness $h_0$, initial pressure $p(x,0)$, initial velocity $u(x,0)$, Coriolis coefficient $\alpha_c$, diastolic pressure $p_D$ (in this model coincident with the external pressure $p_{ext}$ for the equilibrium), outflow pressure $p_{out}$, Windkessel resistance $R_1$, Windkessel resistance $R_2$, Windkessel compliance $C$, instantaneous Young modulus $E_0$, asymptotic Young modulus $E_{\infty}$, viscosity coefficient $\eta$, relaxation time $\tau_r$. In all the tests $\rho = 1060$~kg/m$^3$, $\mu = 0.004$~Pa~s. Concerning numerical parameters, $\CFL = 0.9$ in all the simulations; the number of cells in the domain is $nc = 12$ for tests in the TA and $nc = 6$ for tests in the CCA; for the TA cases 20 cardiac cycles are simulated, corresponding to a final time $t_{end} = 19.10$ s, while for the CCA cases 9 cardiac cycles are simulated, corresponding to a final time $t_{end} = 9.90$ s.}
\label{tab.dataAlastruey}
\end{table}
\subsection{Thoracic aorta benchmark test cases}
\label{section_TC_aorta}
The constant radius upper thoracic aorta test case (cTA) is simulated using a purely elastic wall model to allow comparisons with benchmark available data, from which also the flow rate imposed at the inlet ($q_{in}$) is taken \cite{boileau2015}. Figure~\ref{fig.BA} shows a comparison of the numerical results obtained solving the 1D a-FSI system \eqref{completesyst} with the IMEX scheme against benchmark data. Such data were obtained using a 3D model and six different 1D numerical methods \cite{boileau2015}. It can be observed that, for all the variables, IMEX results are in perfect agreement with 1D benchmark solutions. Larger but still acceptable differences are observed when compared to the 3D benchmark.\\
To account for spatial variations of the properties along the vessel length, the problem with a linearly tapered upper thoracic aorta (tTA) is simulated, following \cite{xiao2014}. All other geometrical and mechanical parameters are unaltered with respect to the baseline model, including inflow and outflow Windkessel total resistance, $R_T = R_1 + R_2$, and compliance. To evaluate the relevance, in terms of damping mechanism, of viscoelastic effects in a large central artery like the aorta, the tTA simulation is run also considering the viscoelastic SLS model. Results are presented in Fig.~\ref{fig.TA} for three different points in the domain (inlet, midpoint and outlet) and compared to the reference elastic solution. It can be noticed that IMEX elastic results totally agree with the 1D elastic benchmark in all the locations. On the other hand, wall viscosity in IMEX viscoelastic results already plays an important role in the TA, being the viscoelastic behaviour of vessels mainly attributed to smooth muscle cells \cite{gow1968,valdez-jasso2009,battista2015}.
\subsection{Common carotid artery benchmark test cases}
\label{section_TC_CCA}
The constant radius common carotid artery test case (cCCA) is simulated using a purely elastic configuration of the wall mechanics to allow comparisons with benchmark available data, from which also the inlet flow rate ($q_{in}$) is taken \cite{boileau2015}. Figure~\ref{fig.BC} shows results obtained solving the 1D a-FSI system \eqref{completesyst} with the IMEX scheme against the benchmark solution. All the IMEX waveforms are almost indistinguishable from the reference ones, confirming the ability of the model to correctly simulate pulse wave hemodynamics in single arterial segments.\\
Also for the CCA, the problem with a linearly tapered radius (tCCA) is performed as presented in \cite{xiao2014}. Designed in the same manner as for the TA, the simulation is executed initially considering a simple elastic behavior of the vessel wall and further taking into account viscoelasticity. In Fig.~\ref{fig.TC} it is possible to observe an excellent correspondence between IMEX elastic results and the 1D elastic benchmark along the whole length of the vessel. All other parameters being equal, the introduction of the viscoelastic model entails a significant damping effect of  pressure waves, associated with a loss of energy of the system. Since the pulse wave is subject to a viscoelastic response along every arterial segment, the damping effect increases toward the periphery of the cardiovascular system: the frequency of the wave increases as the wall viscosity $\eta$, while the equilibrium radius $R_0$ decreases \cite{alastruey2012,alastruey2012a,mynard2015,valdez-jasso2009}, with a relaxation time of the wall that behaves almost like a biological constant \cite{ghigo2016}. It can be verified that this concept is well reproduced by the proposed model when comparing damping effects in the TA (Fig.~\ref{fig.TA}) to those in the CCA (Fig.~\ref{fig.TC}).
\begin{figure}[p!]
\begin{subfigure}{0.33\textwidth}
\centering
\includegraphics[width=0.95\linewidth]{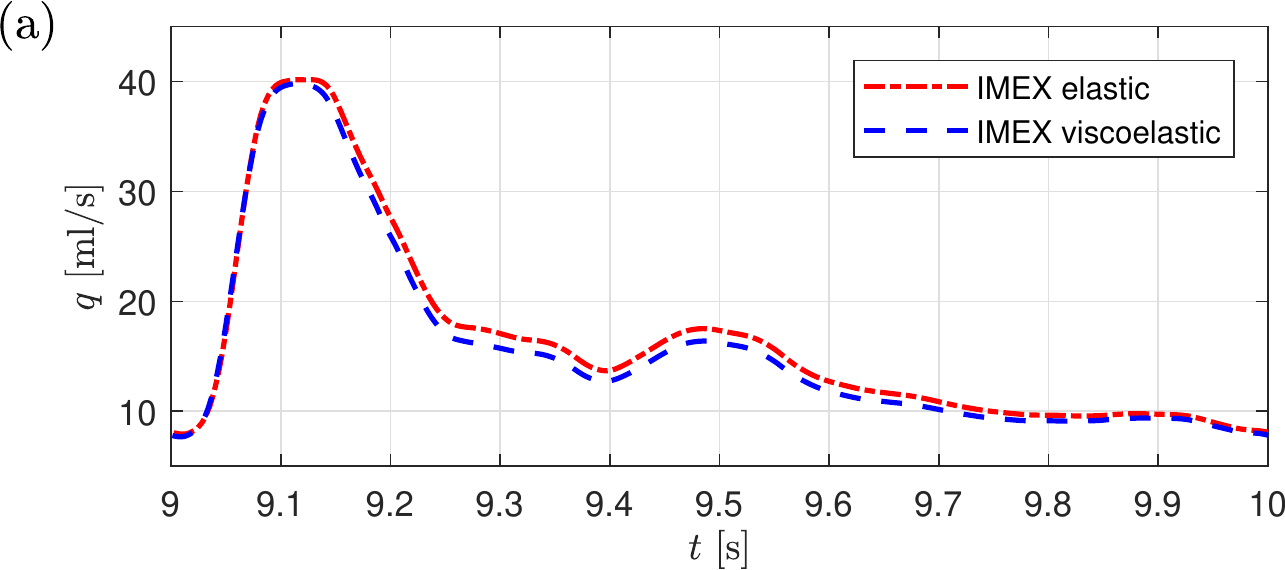}
\label{fig.CarotidNavas_qin}
\end{subfigure}
\begin{subfigure}{0.33\textwidth}
\centering
\includegraphics[width=0.95\linewidth]{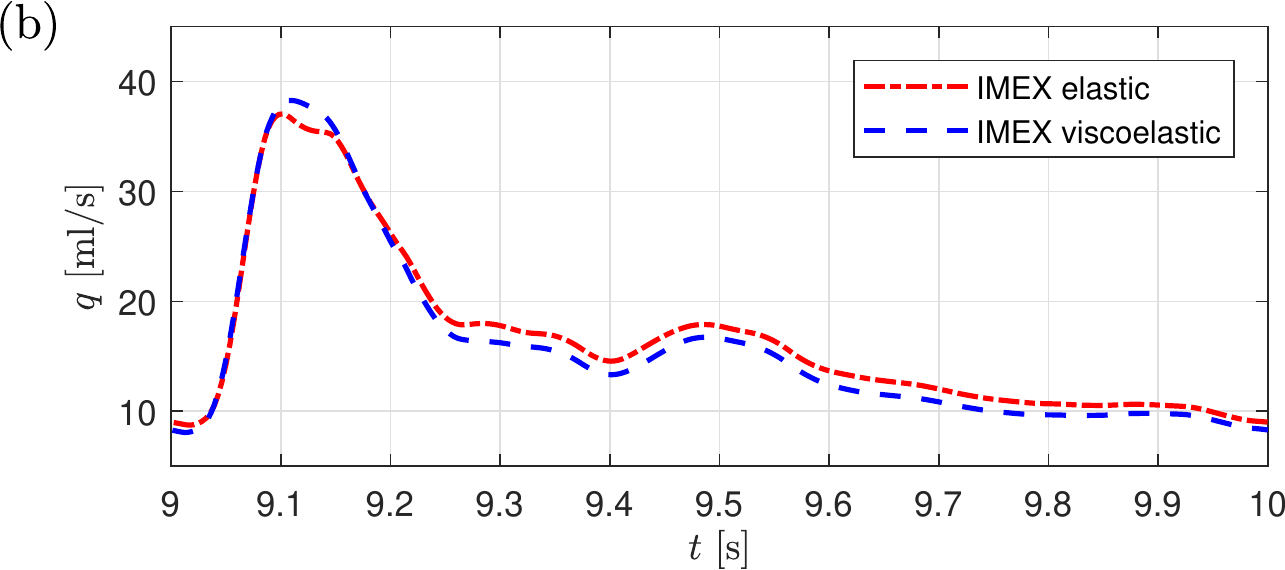}
\label{fig.CarotidNavas_qmid}
\end{subfigure}
\begin{subfigure}{0.33\textwidth}
\centering
\includegraphics[width=0.95\linewidth]{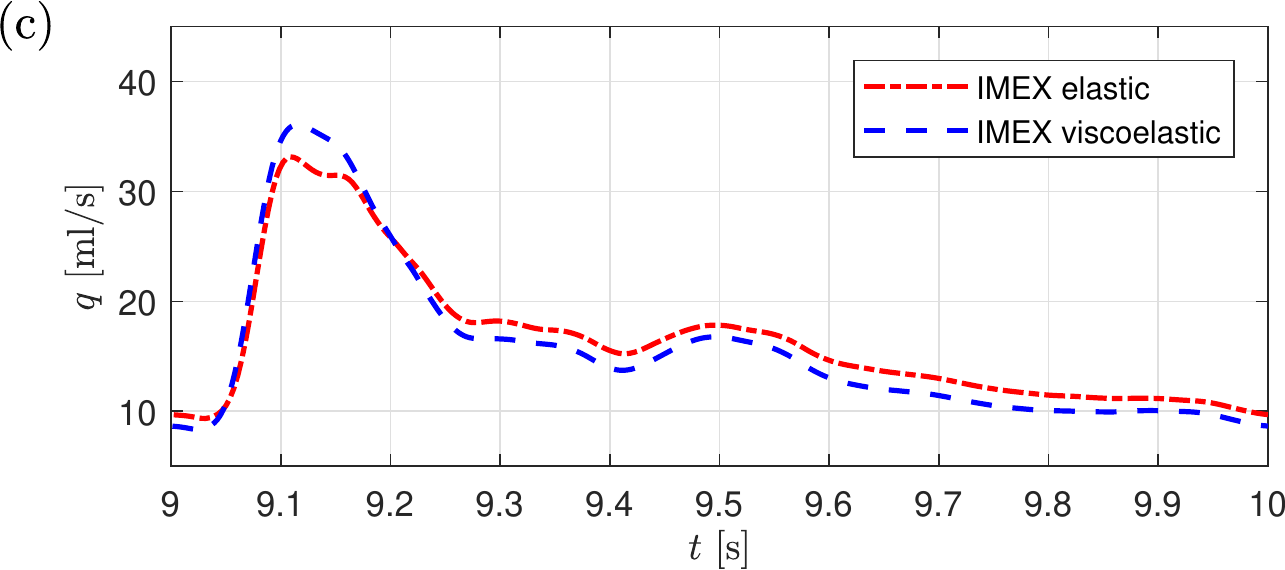}
\label{fig.CarotidNavas_qout}
\end{subfigure}
\begin{subfigure}{0.33\textwidth}
\centering
\includegraphics[width=0.95\linewidth]{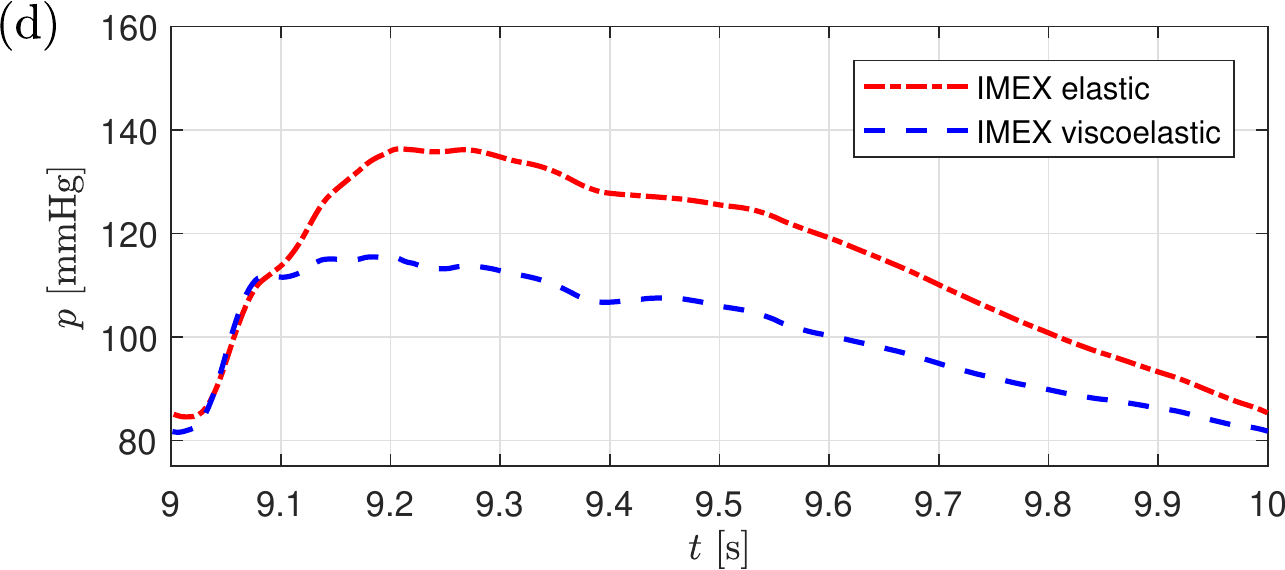}
\label{fig.CarotidNavas_pin}
\vspace*{3mm}
\end{subfigure}
\begin{subfigure}{0.33\textwidth}
\centering
\includegraphics[width=0.95\linewidth]{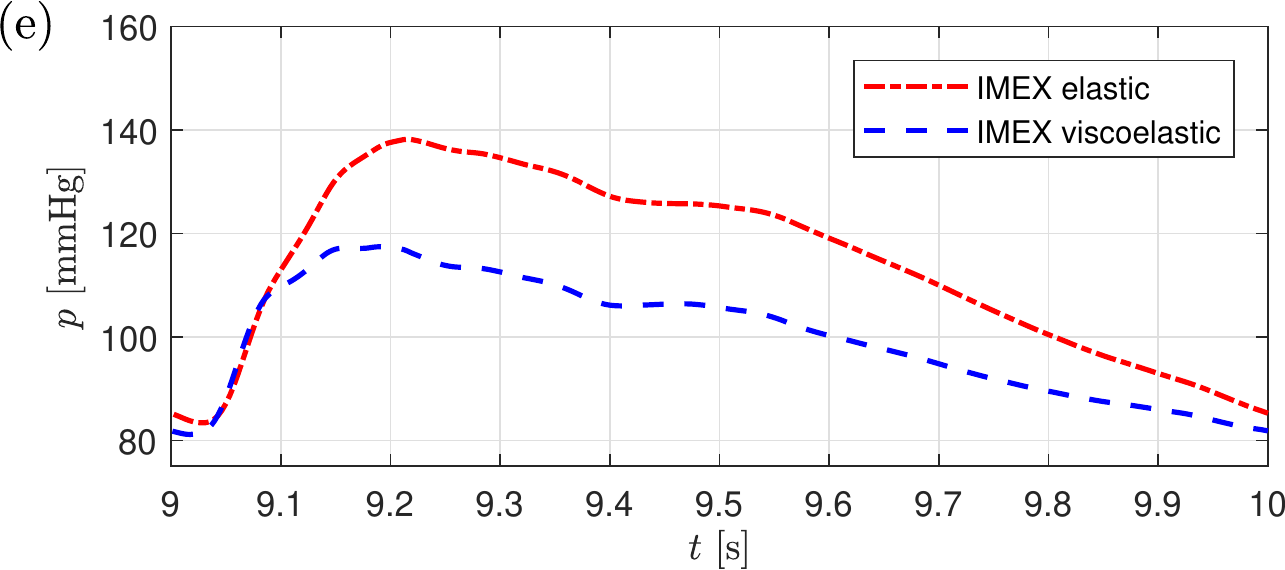}
\label{fig.CarotidNavas_pmid}
\vspace*{3mm}
\end{subfigure}
\begin{subfigure}{0.33\textwidth}
\centering
\includegraphics[width=0.95\linewidth]{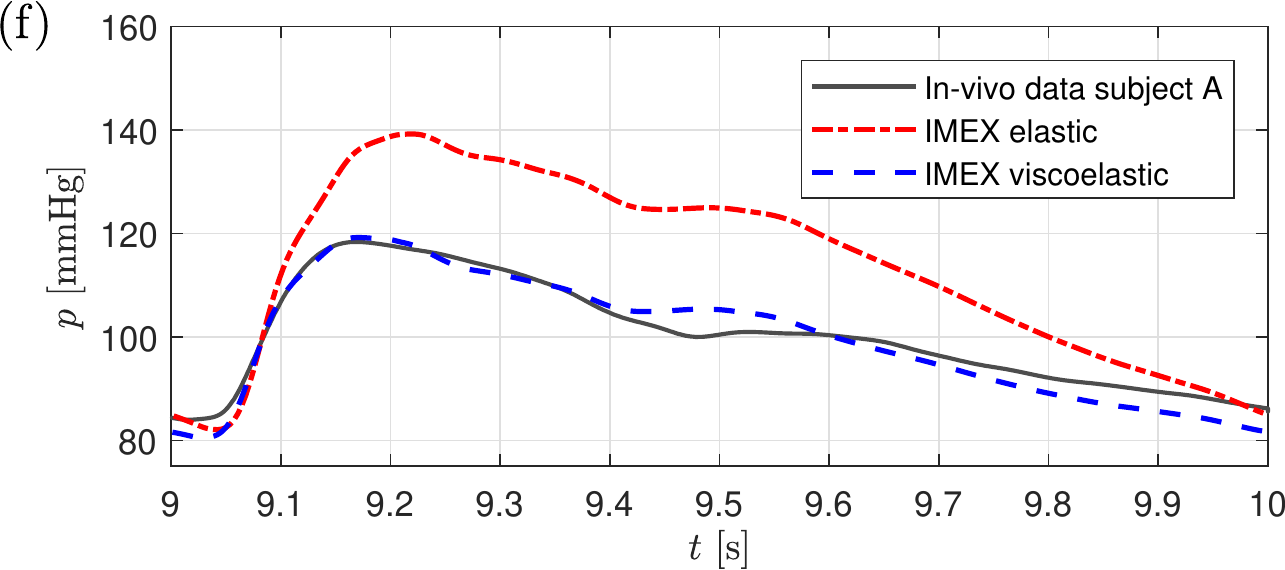}
\label{fig.CarotidNavas_pout}
\vspace*{3mm}
\end{subfigure}

\begin{subfigure}{0.33\textwidth}
\centering
\includegraphics[width=0.95\linewidth]{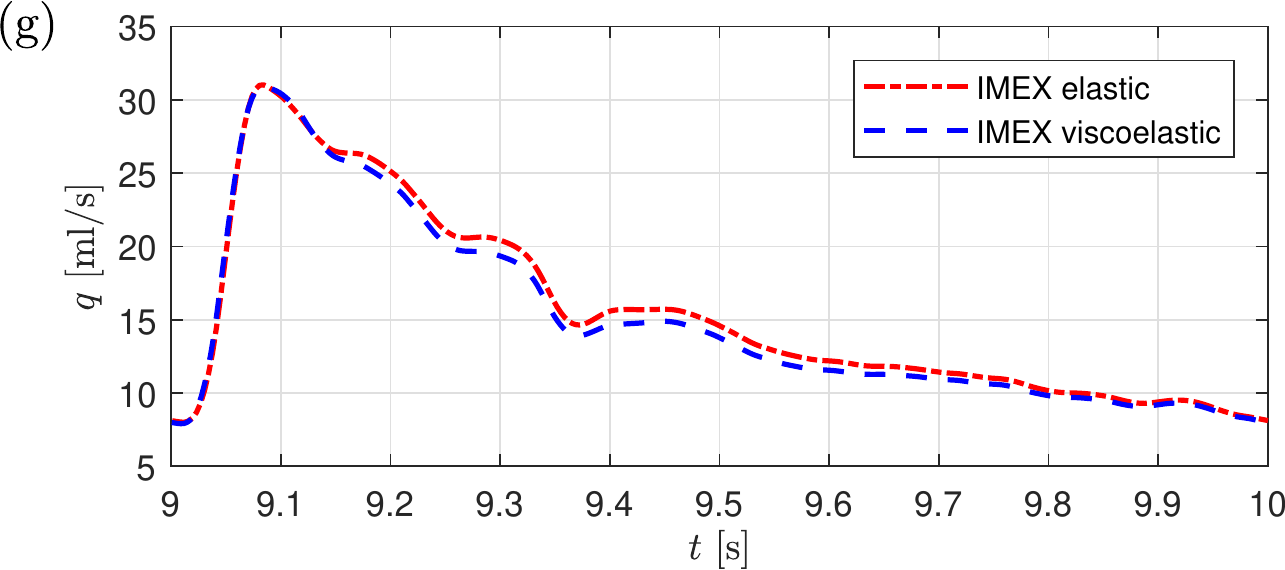}
\label{fig.CarotidMoreno_qin}
\end{subfigure}
\begin{subfigure}{0.33\textwidth}
\centering
\includegraphics[width=0.95\linewidth]{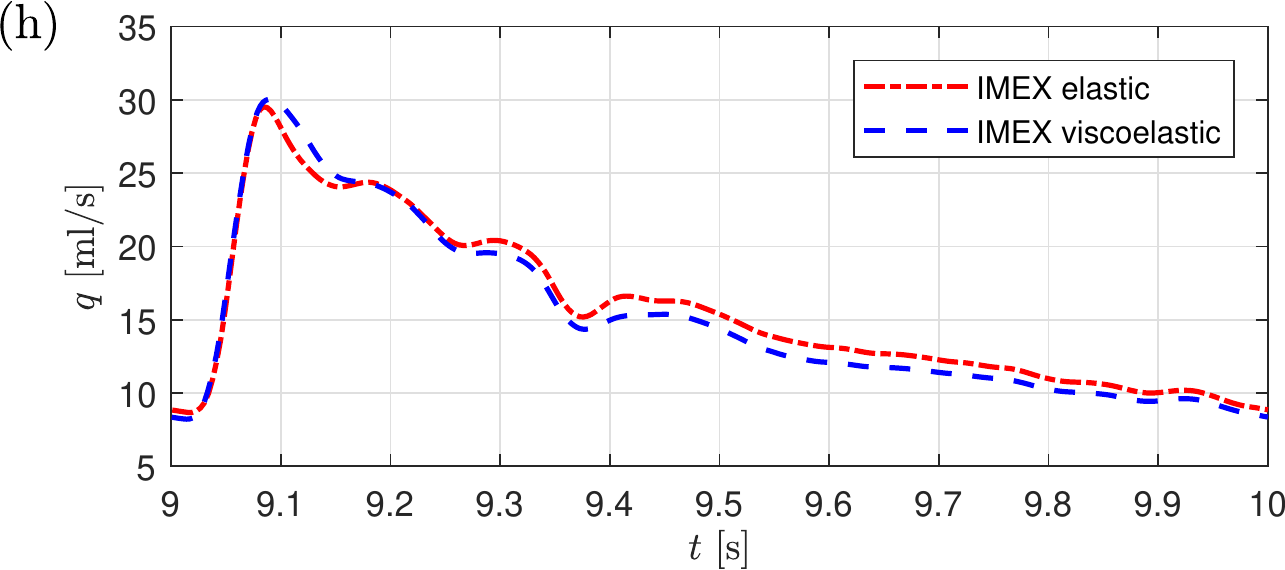}
\label{fig.CarotidMoreno_qmid}
\end{subfigure}
\begin{subfigure}{0.33\textwidth}
\centering
\includegraphics[width=0.95\linewidth]{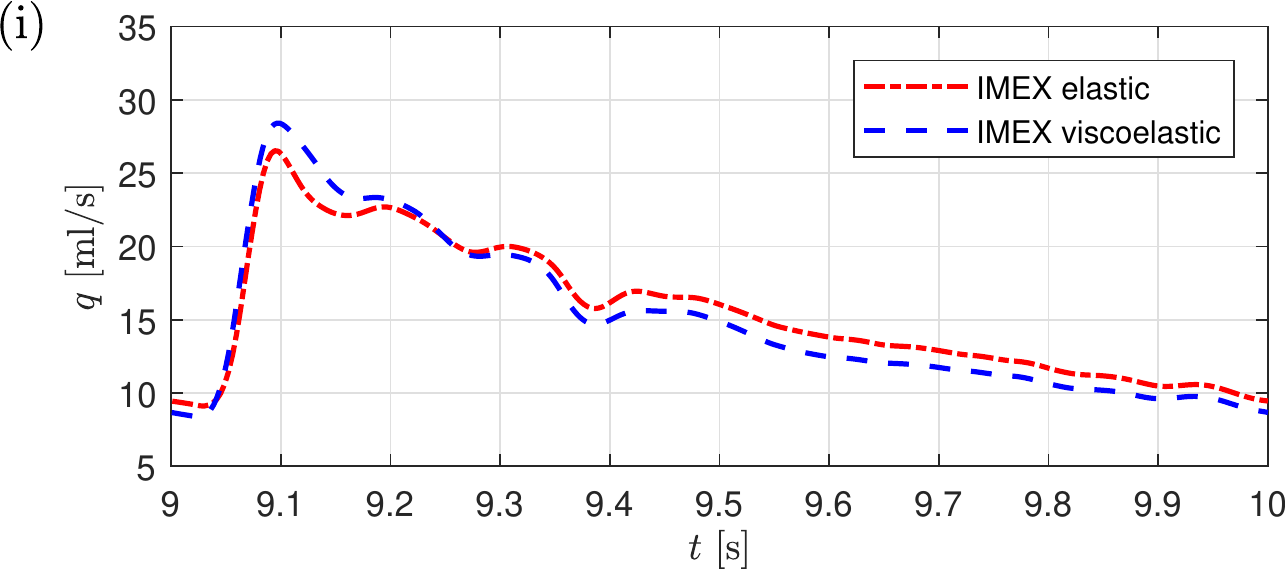}
\label{fig.CarotidMoreno_qout}
\end{subfigure}
\begin{subfigure}{0.33\textwidth}
\centering
\includegraphics[width=0.95\linewidth]{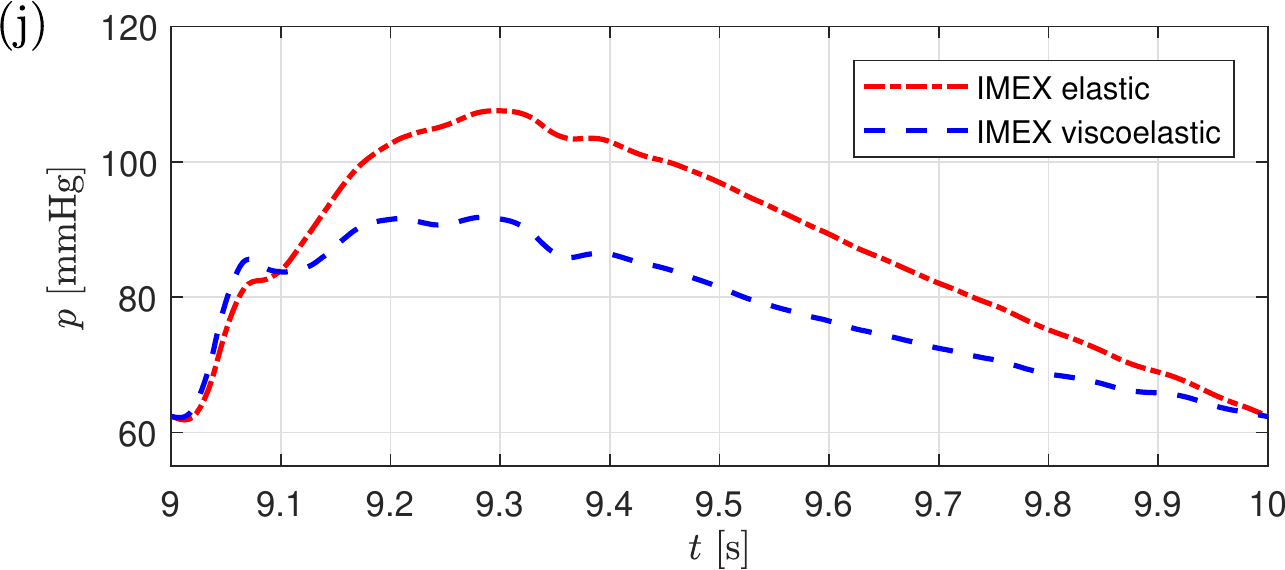}
\label{fig.CarotidMoreno_pin}
\vspace*{3mm}
\end{subfigure}
\begin{subfigure}{0.33\textwidth}
\centering
\includegraphics[width=0.95\linewidth]{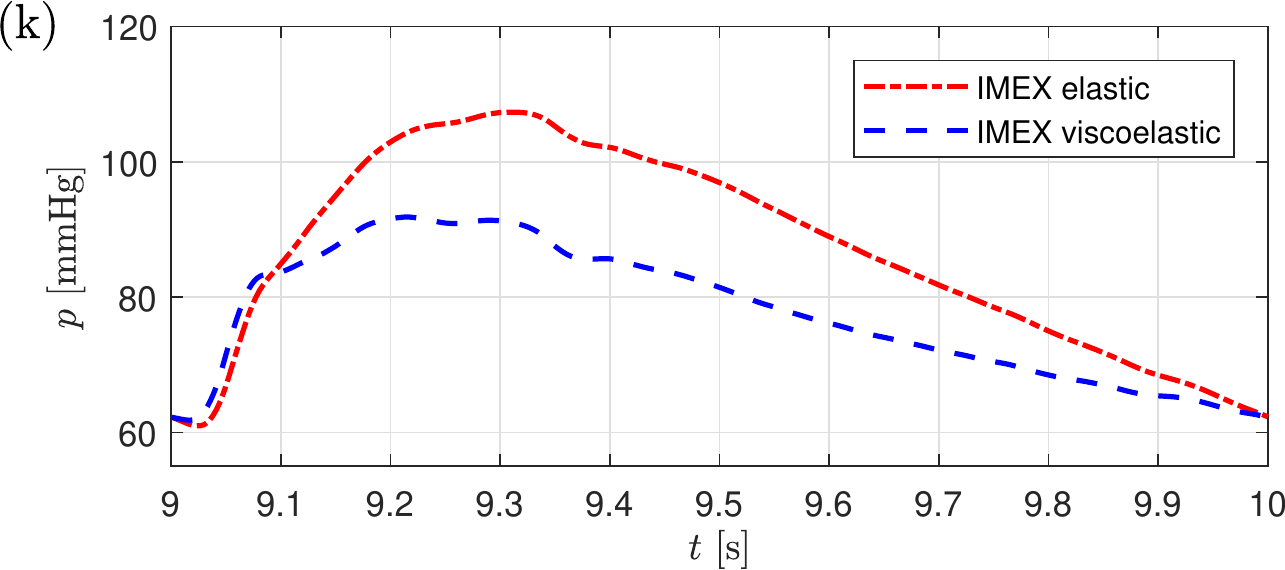}
\label{fig.CarotidMoreno_pmid}
\vspace*{3mm}
\end{subfigure}
\begin{subfigure}{0.33\textwidth}
\centering
\includegraphics[width=0.95\linewidth]{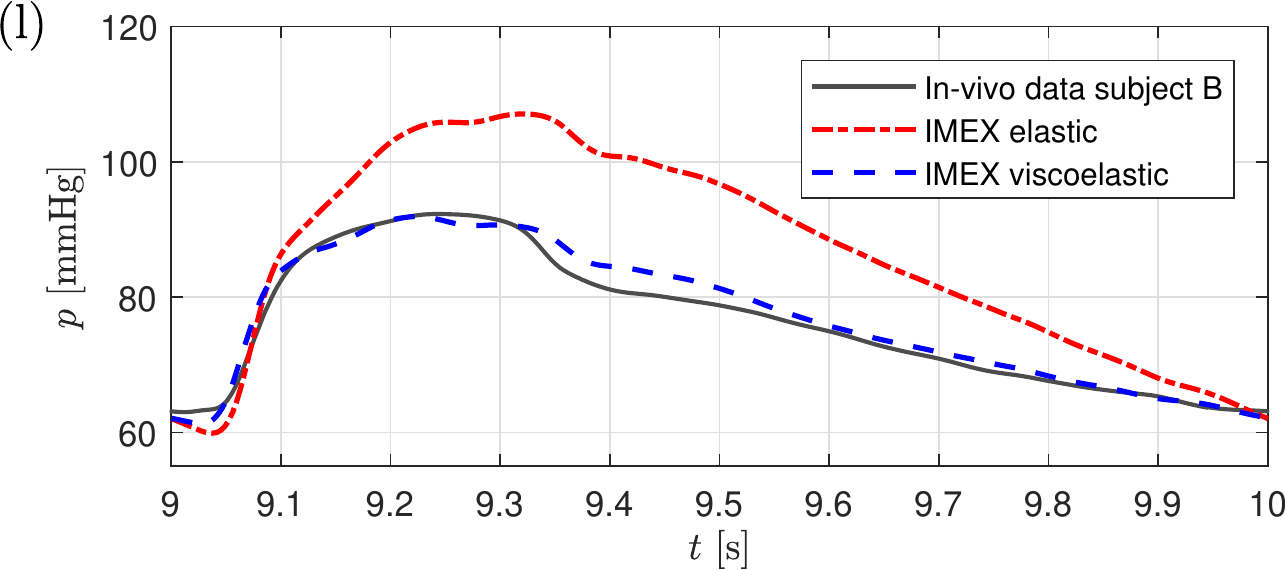}
\label{fig.CarotidMoreno_pout}
\vspace*{3mm}
\end{subfigure}

\begin{subfigure}{0.33\textwidth}
\centering
\includegraphics[width=0.95\linewidth]{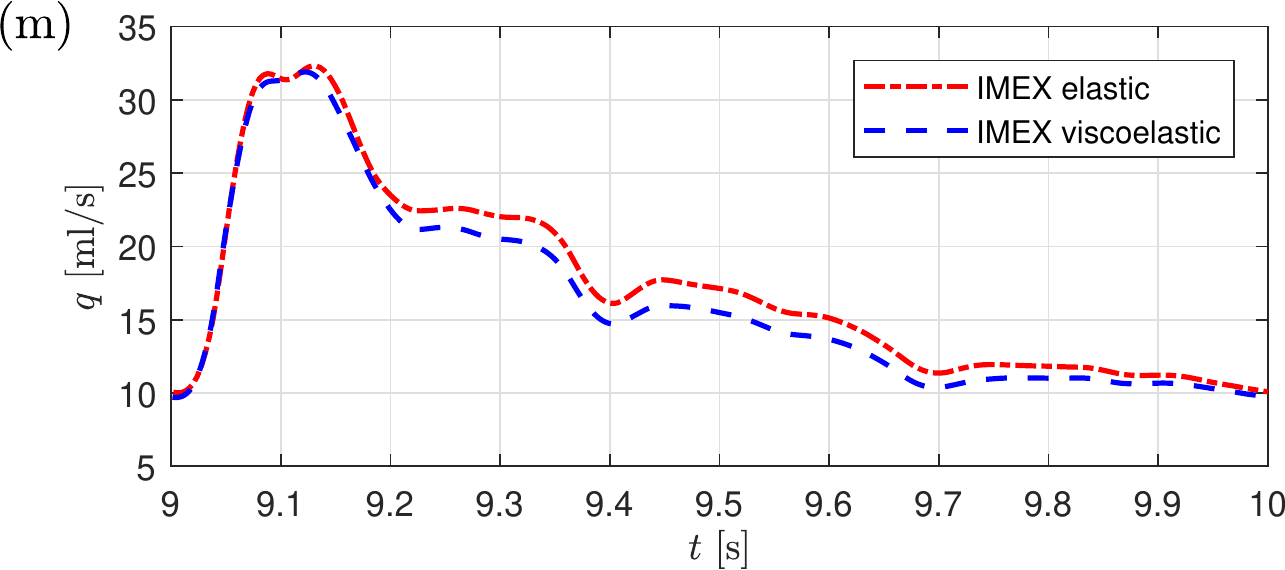}
\label{fig.CarotidRamos_qin}
\end{subfigure}
\begin{subfigure}{0.33\textwidth}
\centering
\includegraphics[width=0.95\linewidth]{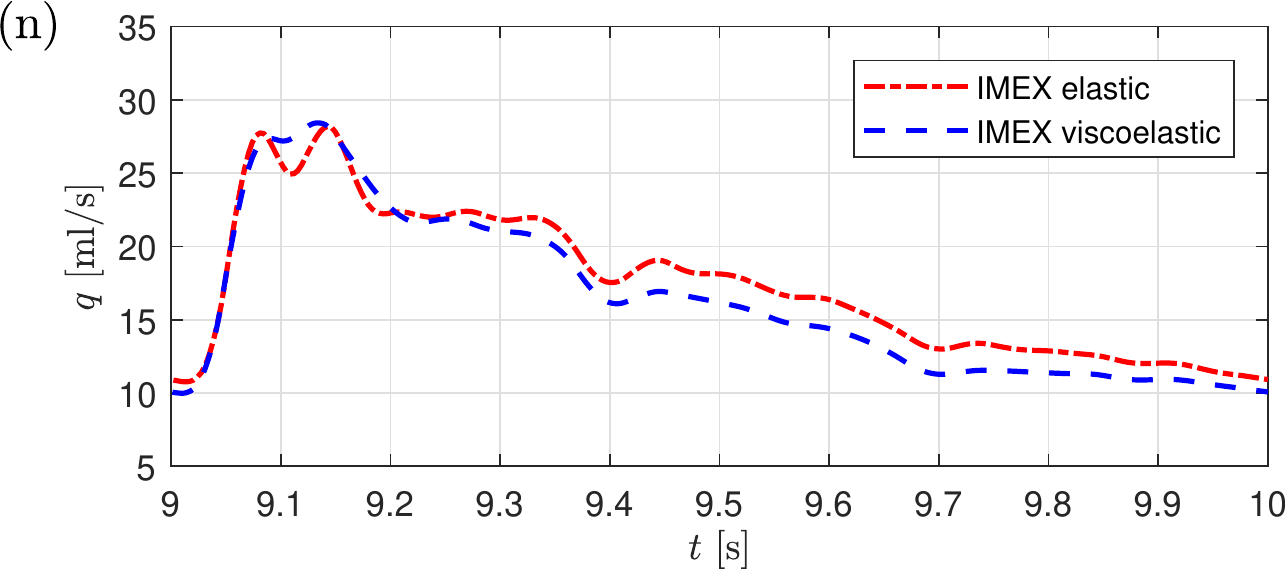}
\label{fig.CarotidRamos_qmid}
\end{subfigure}
\begin{subfigure}{0.33\textwidth}
\centering
\includegraphics[width=0.95\linewidth]{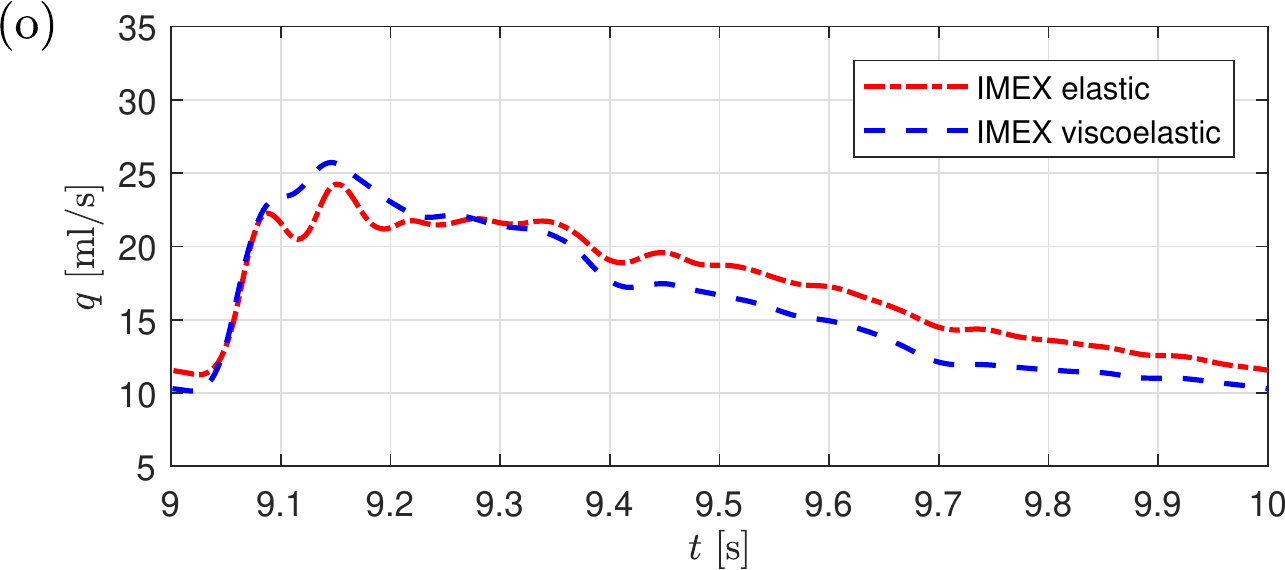}
\label{fig.CarotidRamos_qout}
\end{subfigure}
\begin{subfigure}{0.33\textwidth}
\centering
\includegraphics[width=0.95\linewidth]{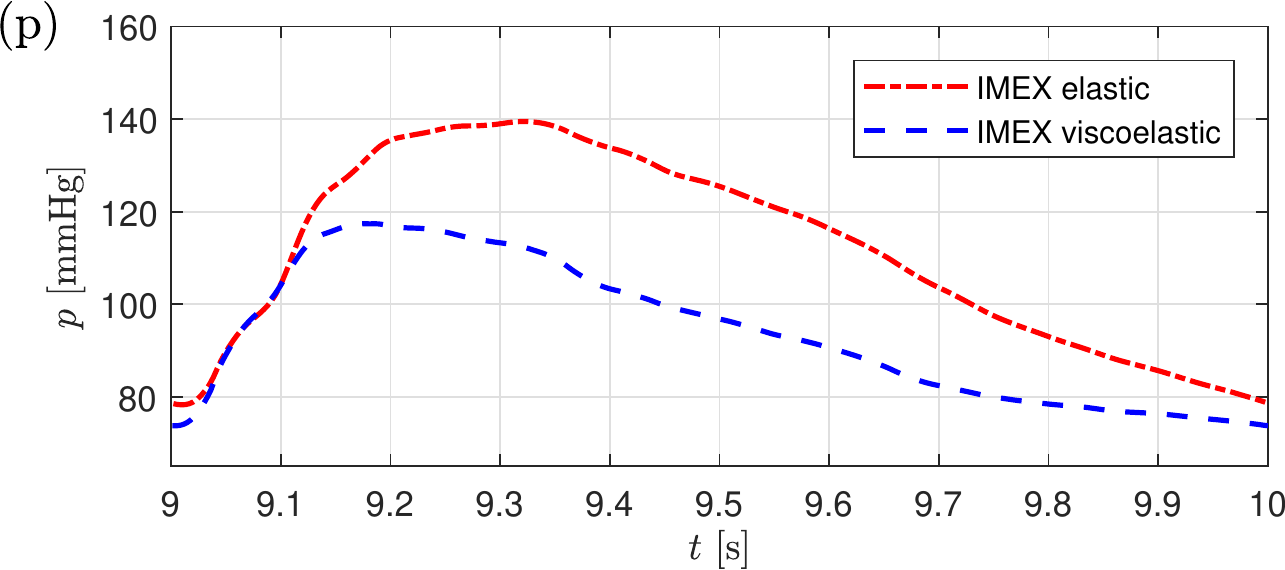}
\label{fig.CarotidRamos_pin}
\vspace*{3mm}
\end{subfigure}
\begin{subfigure}{0.33\textwidth}
\centering
\includegraphics[width=0.95\linewidth]{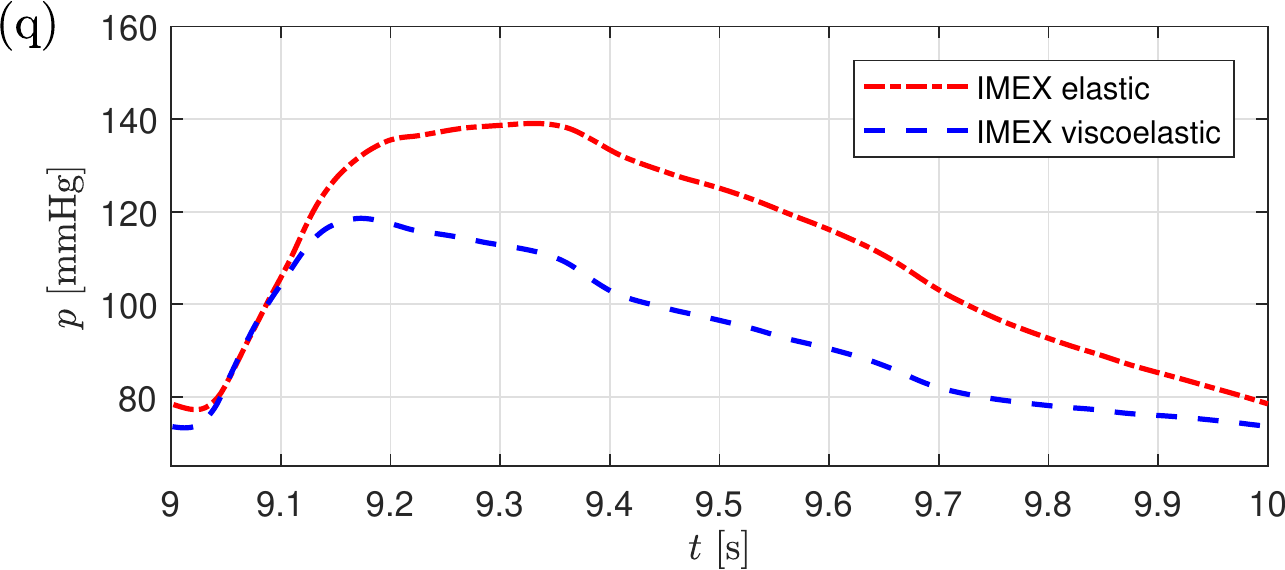}
\label{fig.CarotidRamos_pmid}
\vspace*{3mm}
\end{subfigure}
\begin{subfigure}{0.33\textwidth}
\centering
\includegraphics[width=0.95\linewidth]{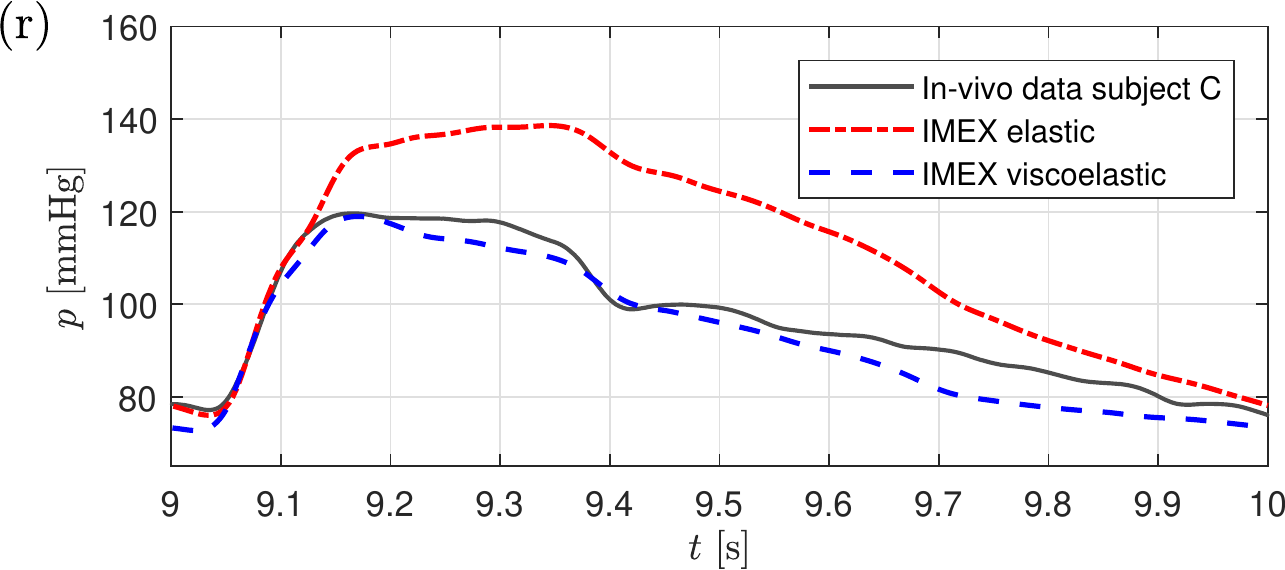}
\label{fig.CarotidRamos_pout}
\vspace*{3mm}
\end{subfigure}

\begin{subfigure}{0.33\textwidth}
\centering
\includegraphics[width=0.95\linewidth]{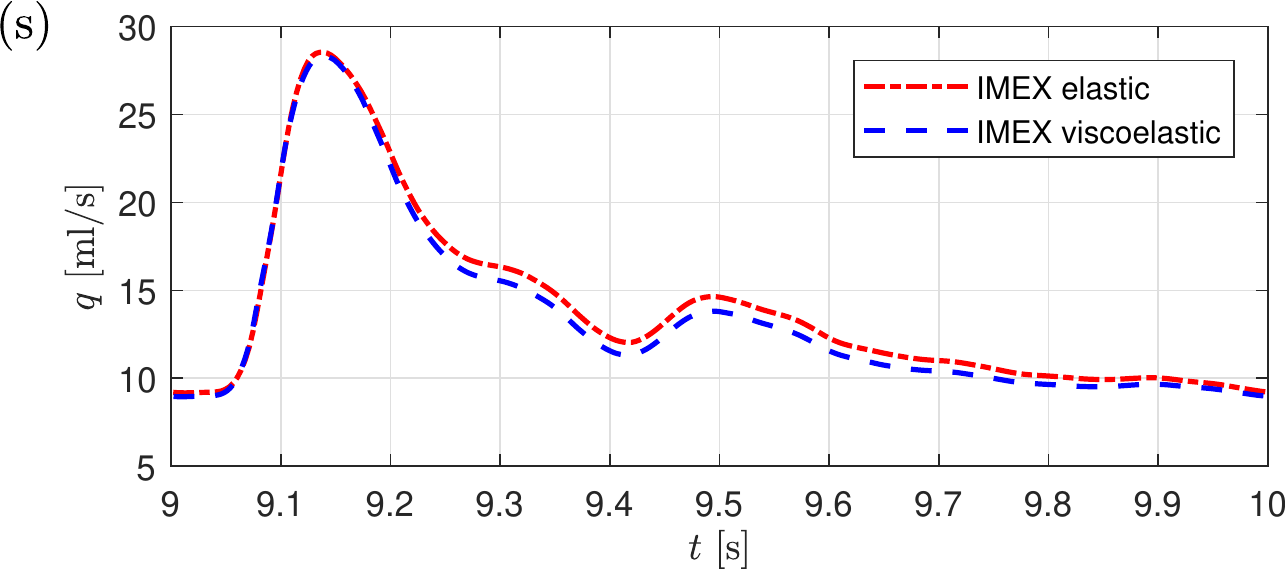}
\label{fig.CarotidBertaglia_qin}
\end{subfigure}
\begin{subfigure}{0.33\textwidth}
\centering
\includegraphics[width=0.95\linewidth]{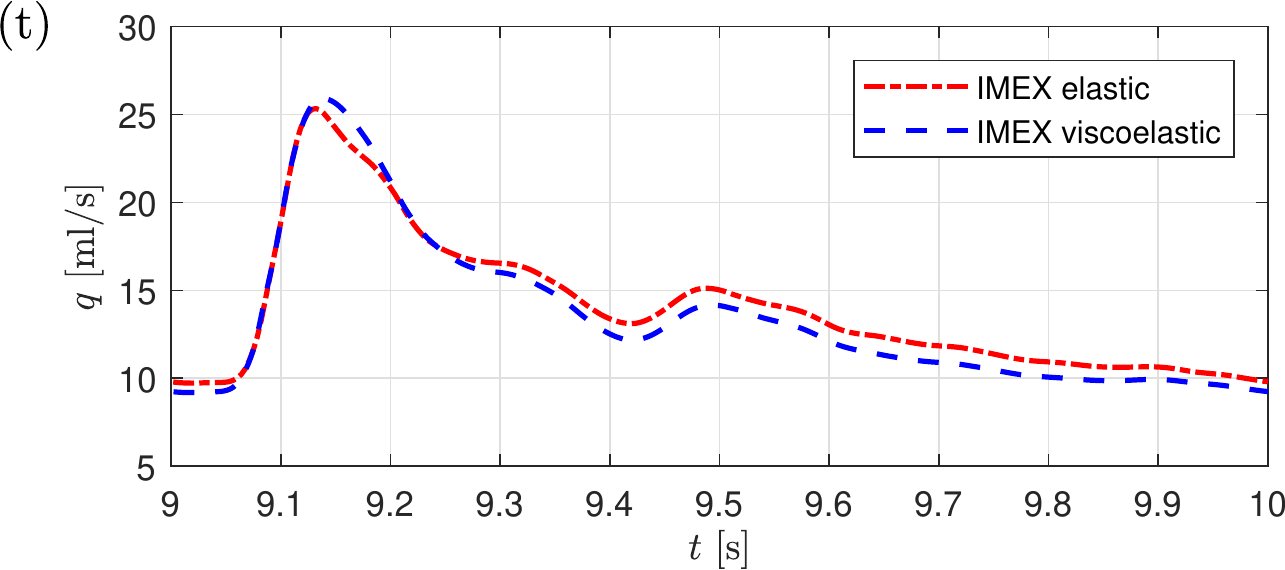}
\label{fig.CarotidBertaglia_qmid}
\end{subfigure}
\begin{subfigure}{0.33\textwidth}
\centering
\includegraphics[width=0.95\linewidth]{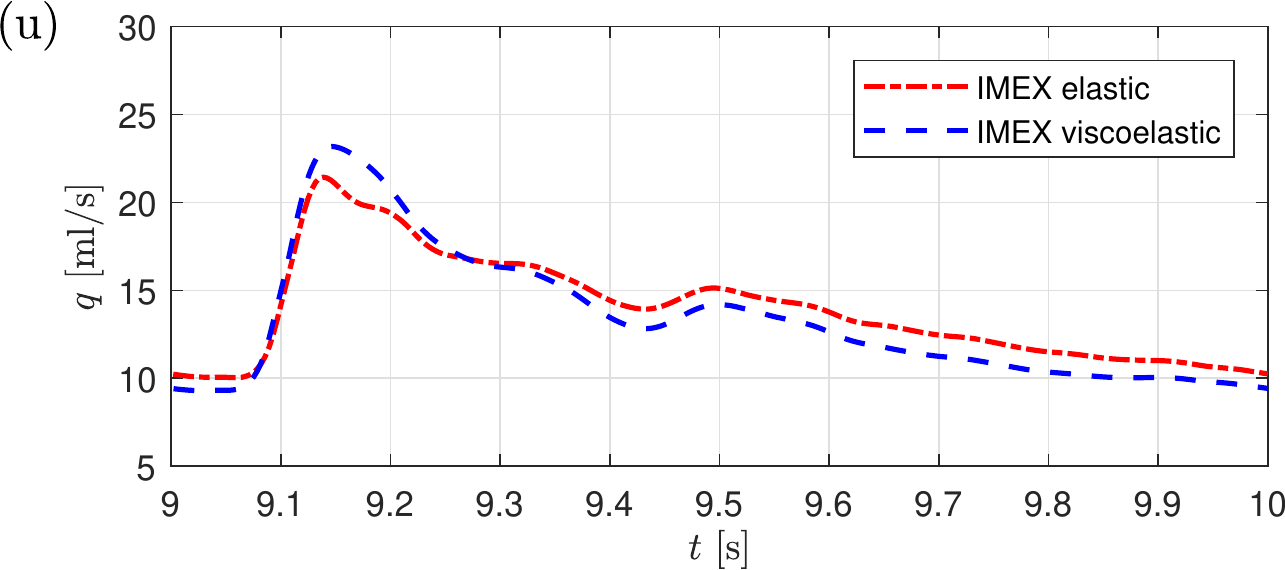}
\label{fig.CarotidBertaglia_qout}
\end{subfigure}
\begin{subfigure}{0.33\textwidth}
\centering
\includegraphics[width=0.95\linewidth]{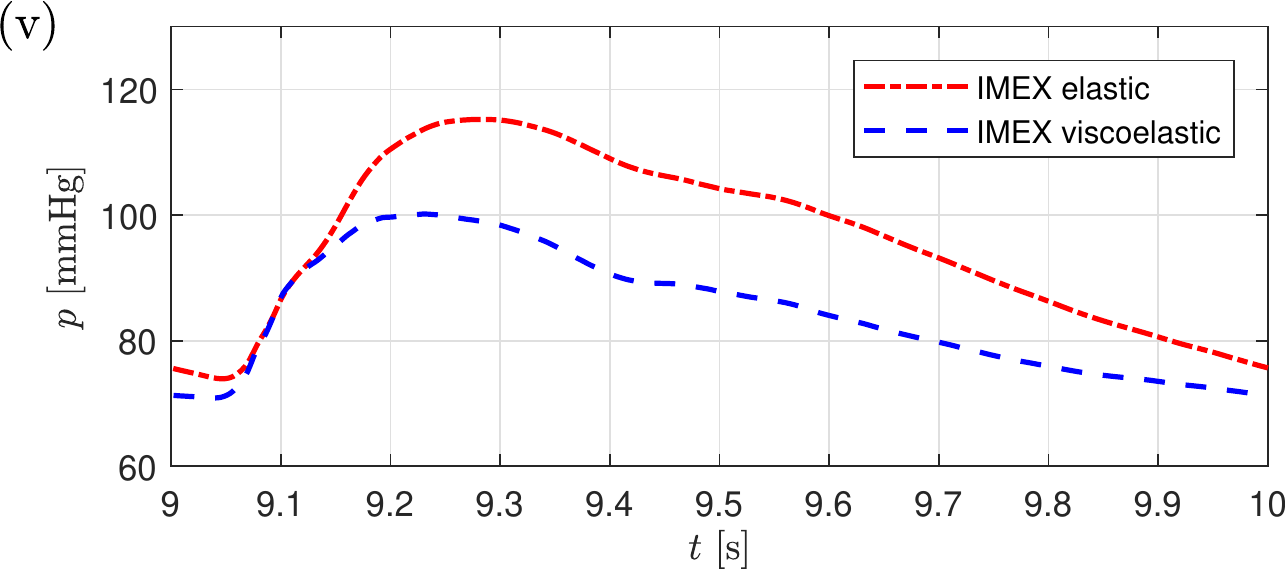}
\label{fig.CarotidBertaglia_pin}
\end{subfigure}
\begin{subfigure}{0.33\textwidth}
\centering
\includegraphics[width=0.95\linewidth]{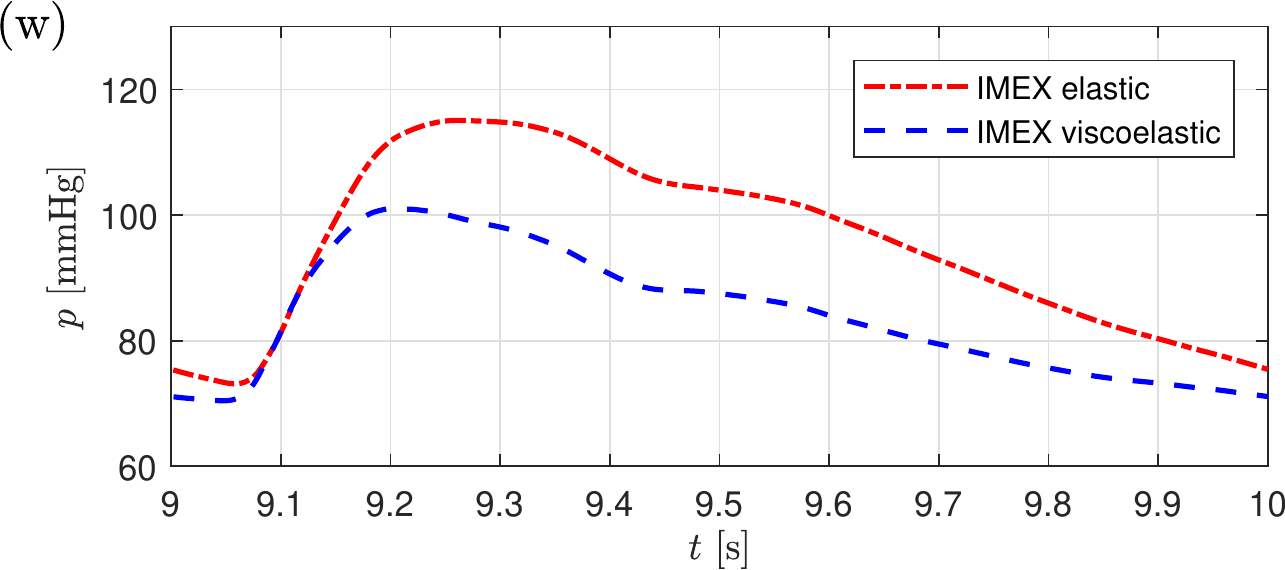}
\label{fig.CarotidBertaglia_pmid}
\end{subfigure}
\begin{subfigure}{0.33\textwidth}
\centering
\includegraphics[width=0.95\linewidth]{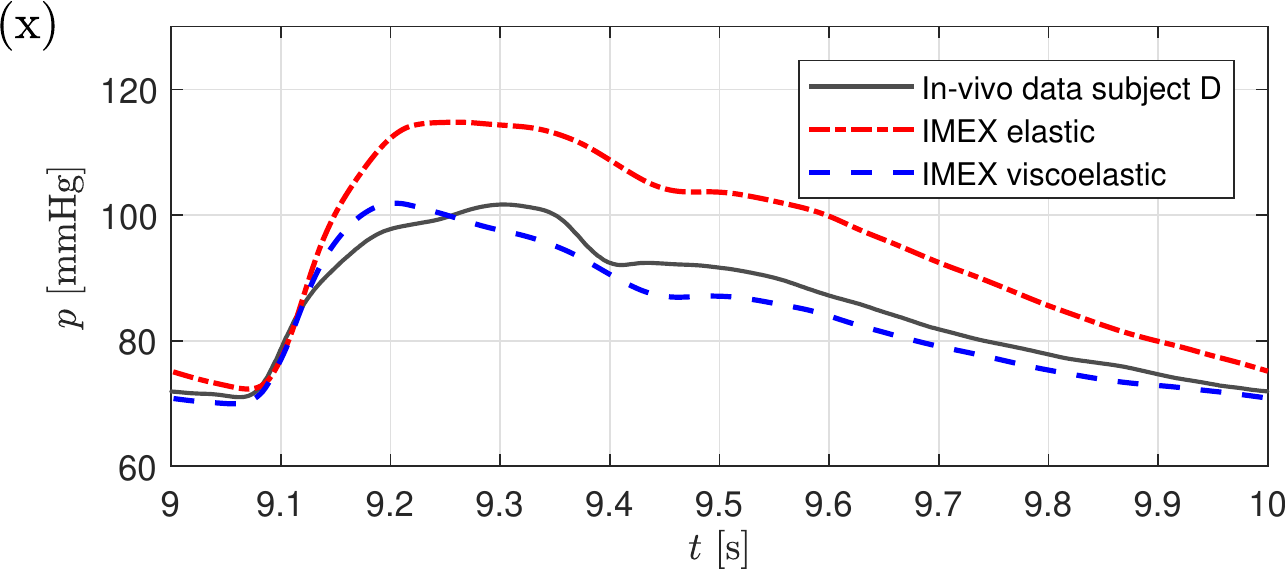}
\label{fig.CarotidBertaglia_pout}
\end{subfigure}
\caption{Common carotid artery (CCA) cases with in-vivo data. Results obtained solving the a-FSI system with the IMEX scheme with elastic and viscoelastic tube law for 4 different subjects. First (flow rate) and second (pressure) rows related to subject A; third (flow rate) and forth (pressure) rows related to subject B; fifth (flow rate) and sixth (pressure) rows related to subject C; seventh (flow rate) and eight (pressure) rows related to subject D. First column shows results in the first cell of the domain, second column shows results in the central cell of the domain, third column shows results in the last cell of the domain. Inlet velocity waveform obtained for each subject from Doppler measurements. Computed pressure obtained in the last cell of the domain compared to pressure waveforms measured for each subject with the PulsePen tonometer.}
\label{fig.carotid_invivo}
\end{figure}
\begin{figure}[h!]
\begin{subfigure}{0.33\textwidth}
\centering
\includegraphics[width=0.95\linewidth]{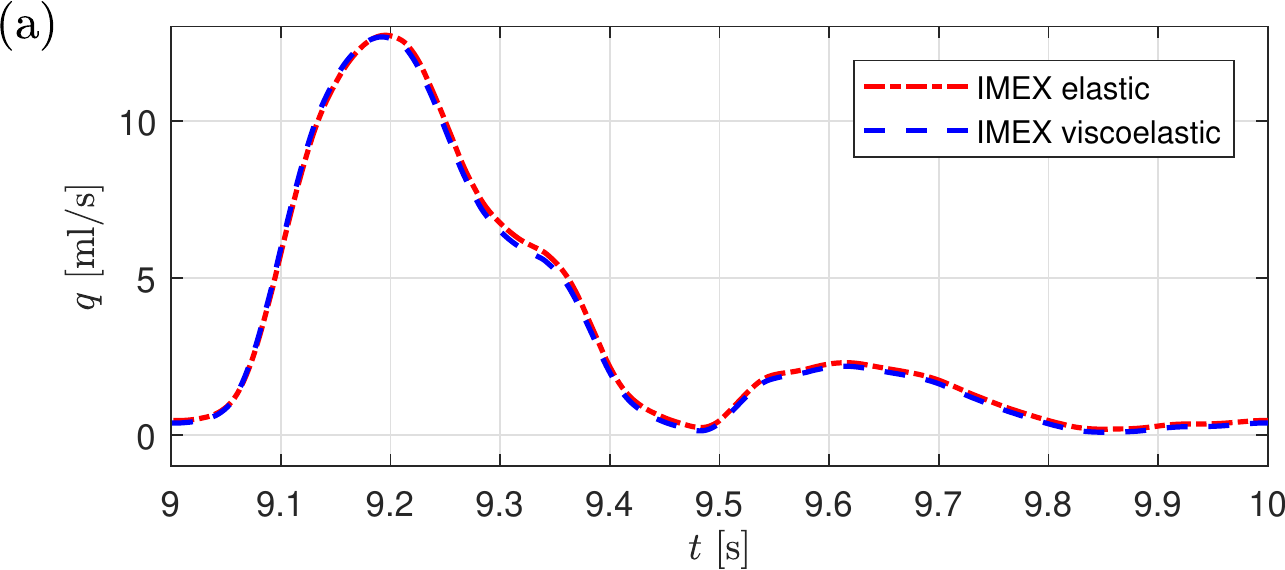}
\label{fig.FemoralSubj1_qin}
\end{subfigure}
\begin{subfigure}{0.33\textwidth}
\centering
\includegraphics[width=0.95\linewidth]{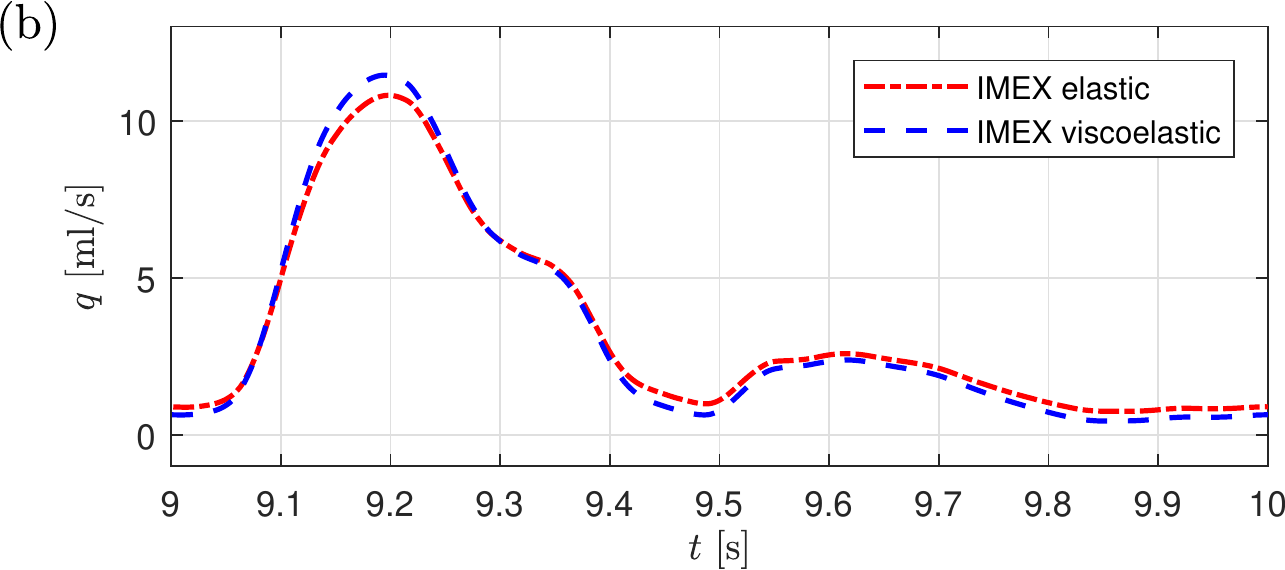}
\label{fig.FemoralSubj1_qmid}
\end{subfigure}
\begin{subfigure}{0.33\textwidth}
\centering
\includegraphics[width=0.95\linewidth]{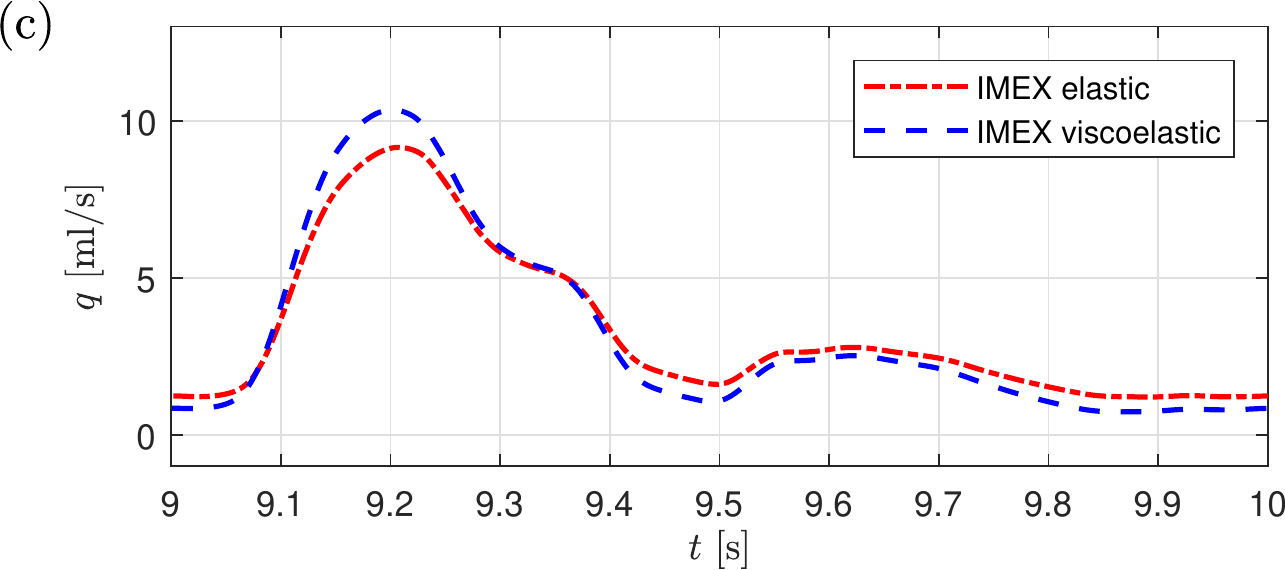}
\label{fig.FemoralSubj1_qout}
\end{subfigure}
\begin{subfigure}{0.33\textwidth}
\centering
\includegraphics[width=0.95\linewidth]{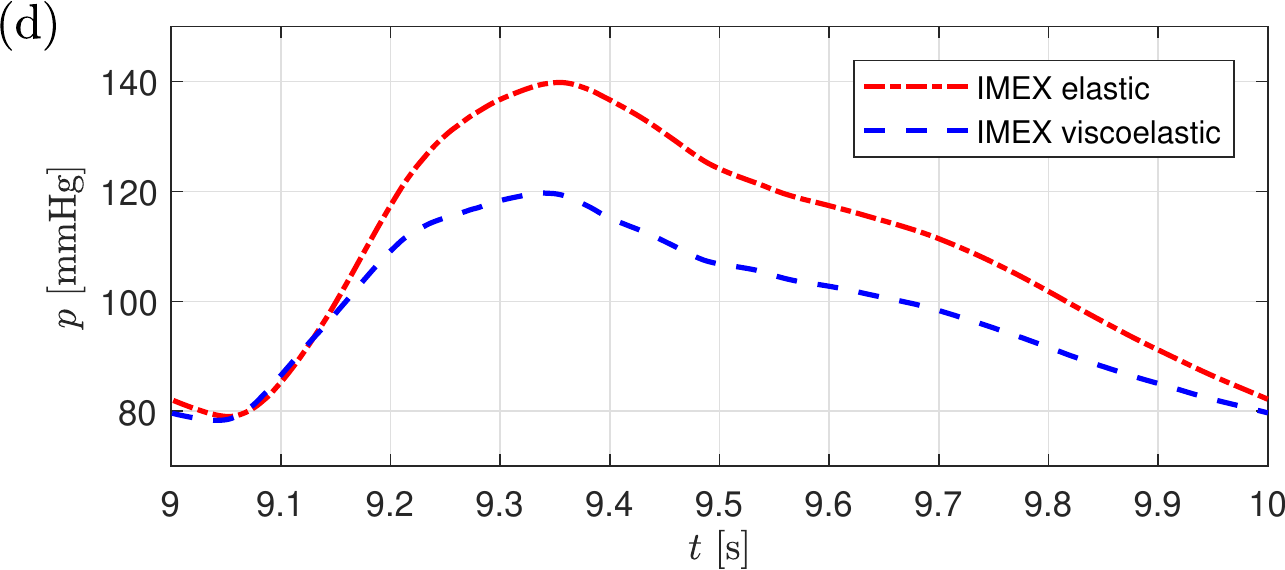}
\label{fig.FemoralSubj1_pin}
\vspace*{3mm}
\end{subfigure}
\begin{subfigure}{0.33\textwidth}
\centering
\includegraphics[width=0.95\linewidth]{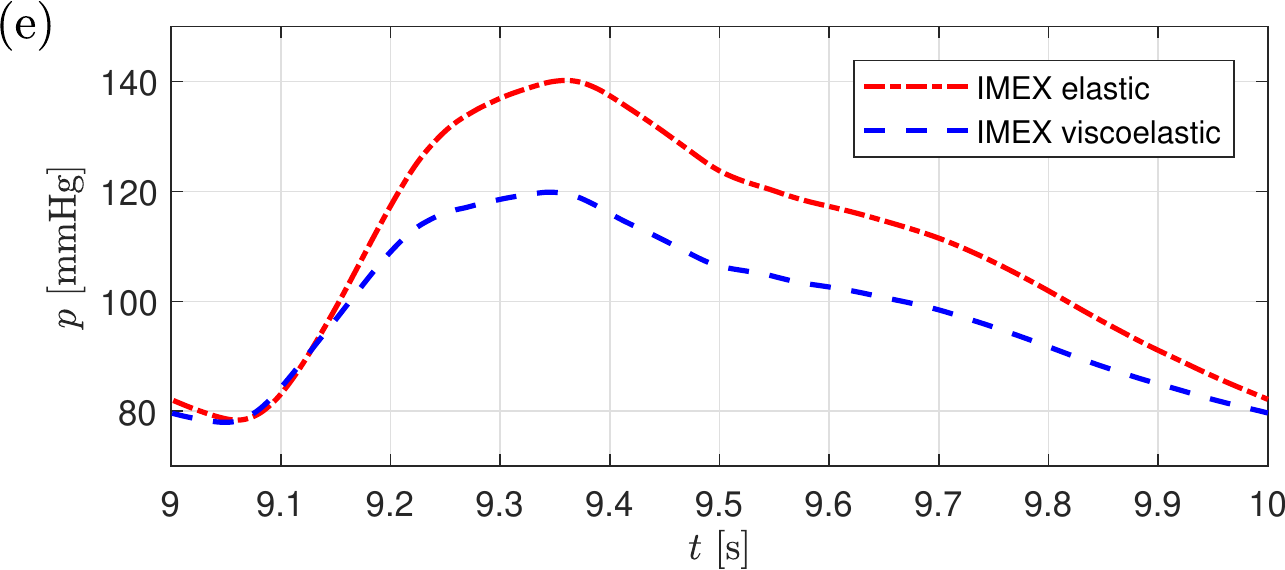}
\label{fig.FemoralSubj1_pmid}
\vspace*{3mm}
\end{subfigure}
\begin{subfigure}{0.33\textwidth}
\centering
\includegraphics[width=0.95\linewidth]{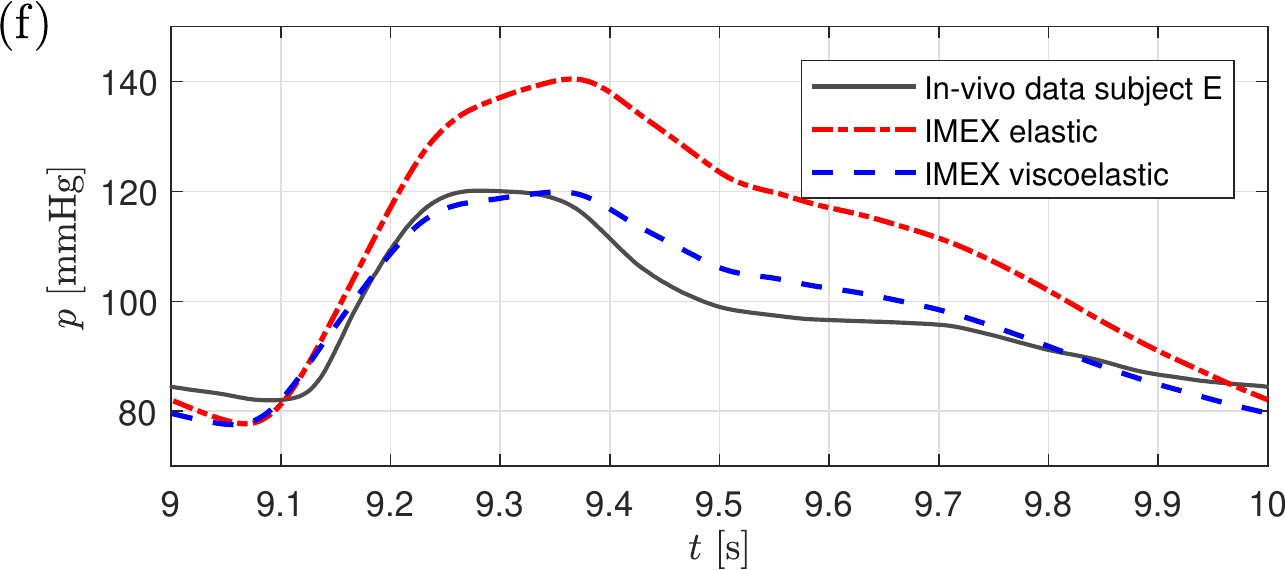}
\label{fig.FemoralSubj1_pout}
\vspace*{3mm}
\end{subfigure}

\begin{subfigure}{0.33\textwidth}
\centering
\includegraphics[width=0.95\linewidth]{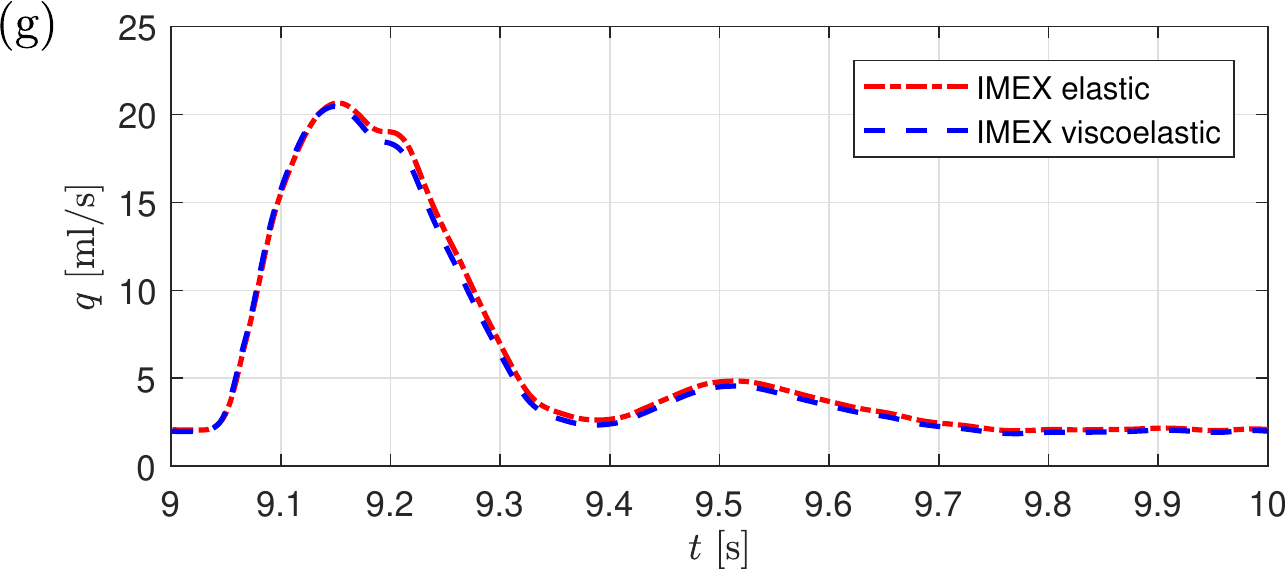}
\label{fig.FemoralSubj2_qin}
\end{subfigure}
\begin{subfigure}{0.33\textwidth}
\centering
\includegraphics[width=0.95\linewidth]{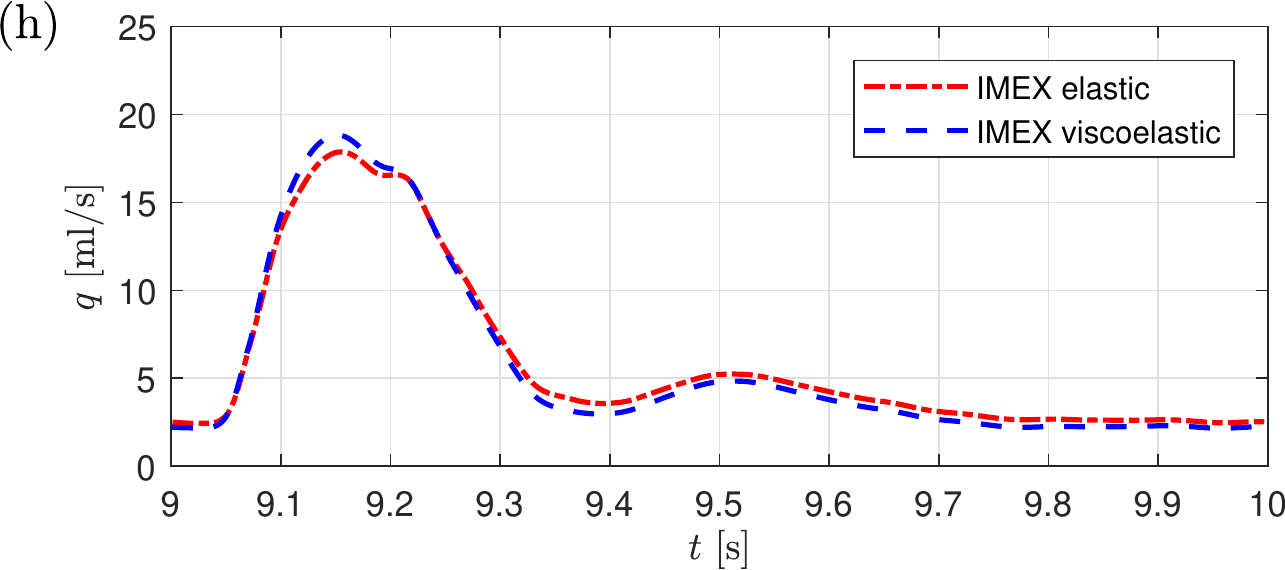}
\label{fig.FemoralSubj2_qmid}
\end{subfigure}
\begin{subfigure}{0.33\textwidth}
\centering
\includegraphics[width=0.95\linewidth]{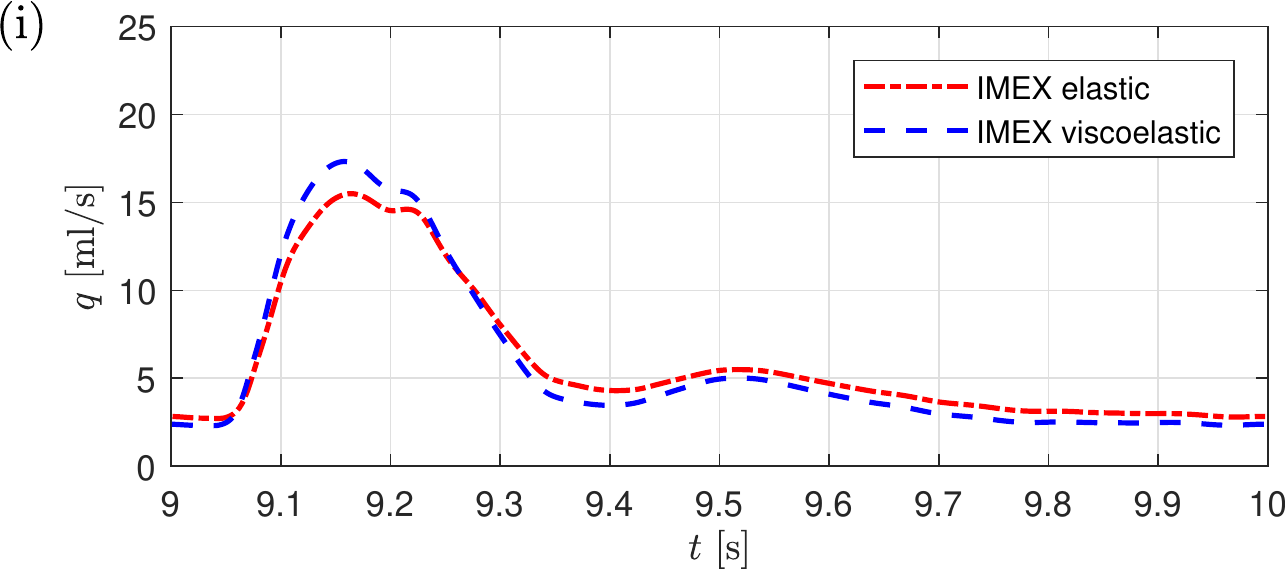}
\label{fig.FemoralSubj2_qout}
\end{subfigure}
\begin{subfigure}{0.33\textwidth}
\centering
\includegraphics[width=0.95\linewidth]{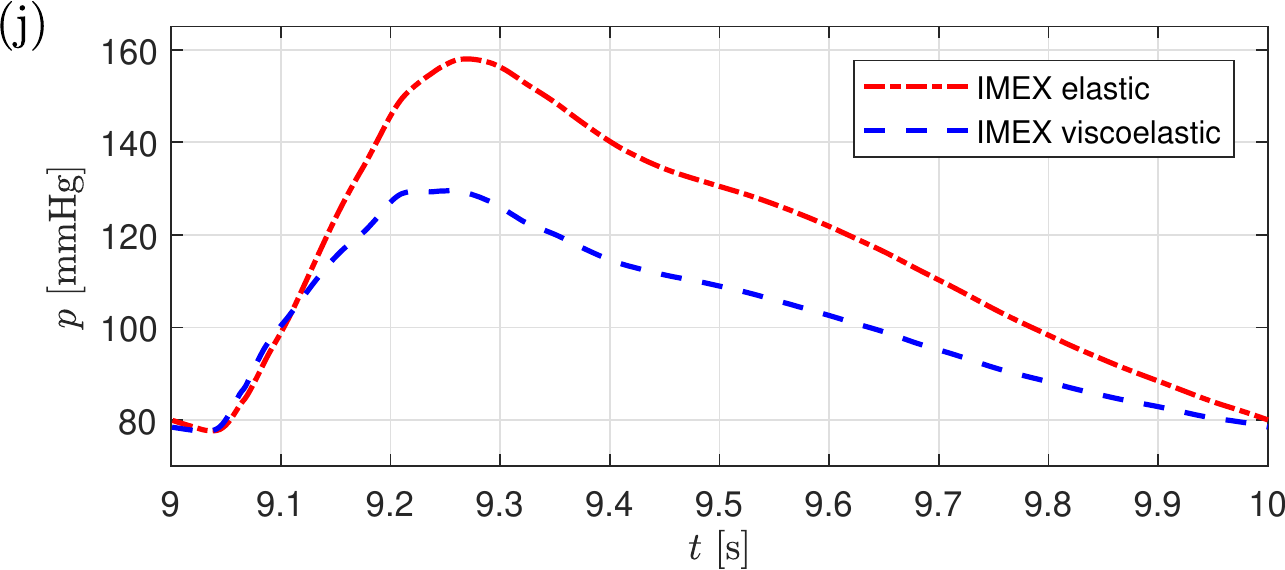}
\label{fig.FemoralSubj2_pin}
\end{subfigure}
\begin{subfigure}{0.33\textwidth}
\centering
\includegraphics[width=0.95\linewidth]{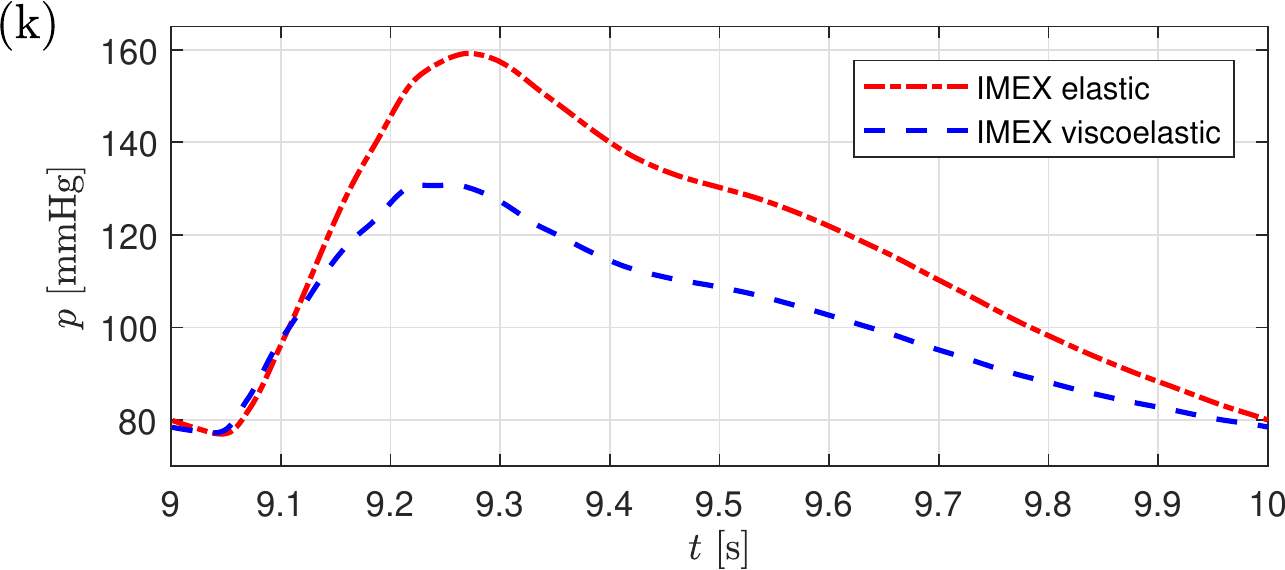}
\label{fig.FemoralSubj2_pmid}
\end{subfigure}
\begin{subfigure}{0.33\textwidth}
\centering
\includegraphics[width=0.95\linewidth]{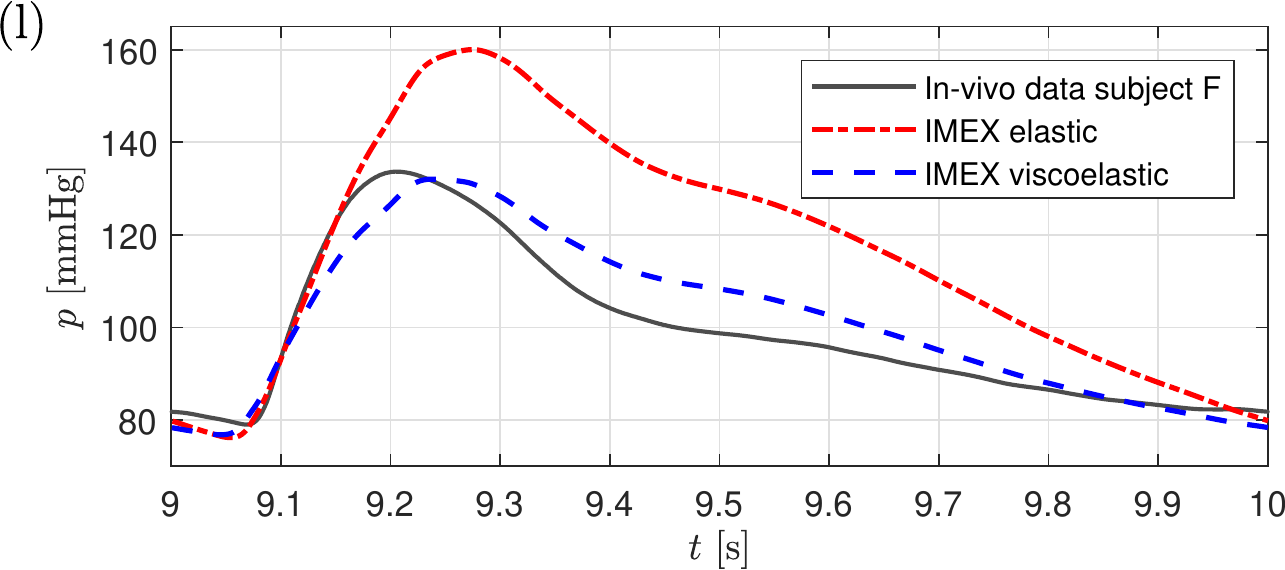}
\label{fig.FemoralSubj2_pout}
\end{subfigure}
\caption{Common femoral artery (CFA) cases with in-vivo data. Results obtained solving the a-FSI system with the IMEX scheme with elastic and viscoelastic tube law for 2 different subjects. First (flow rate) and second (pressure) rows related to subject E; third (flow rate) and forth (pressure) rows related to subject F. First column shows results in the first cell of the domain, second column shows results in the central cell of the domain, third column shows results in the last cell of the domain. Inlet velocity waveform obtained for each subject from Doppler measurements. Computed pressure obtained in the last cell of the domain compared to pressure waveforms measured for each subject with the PulsePen tonometer.}
\label{fig.femoral_invivo}
\end{figure}
\begin{figure}[h!]
\begin{subfigure}{0.49\textwidth}
\centering
\includegraphics[width=0.95\linewidth]{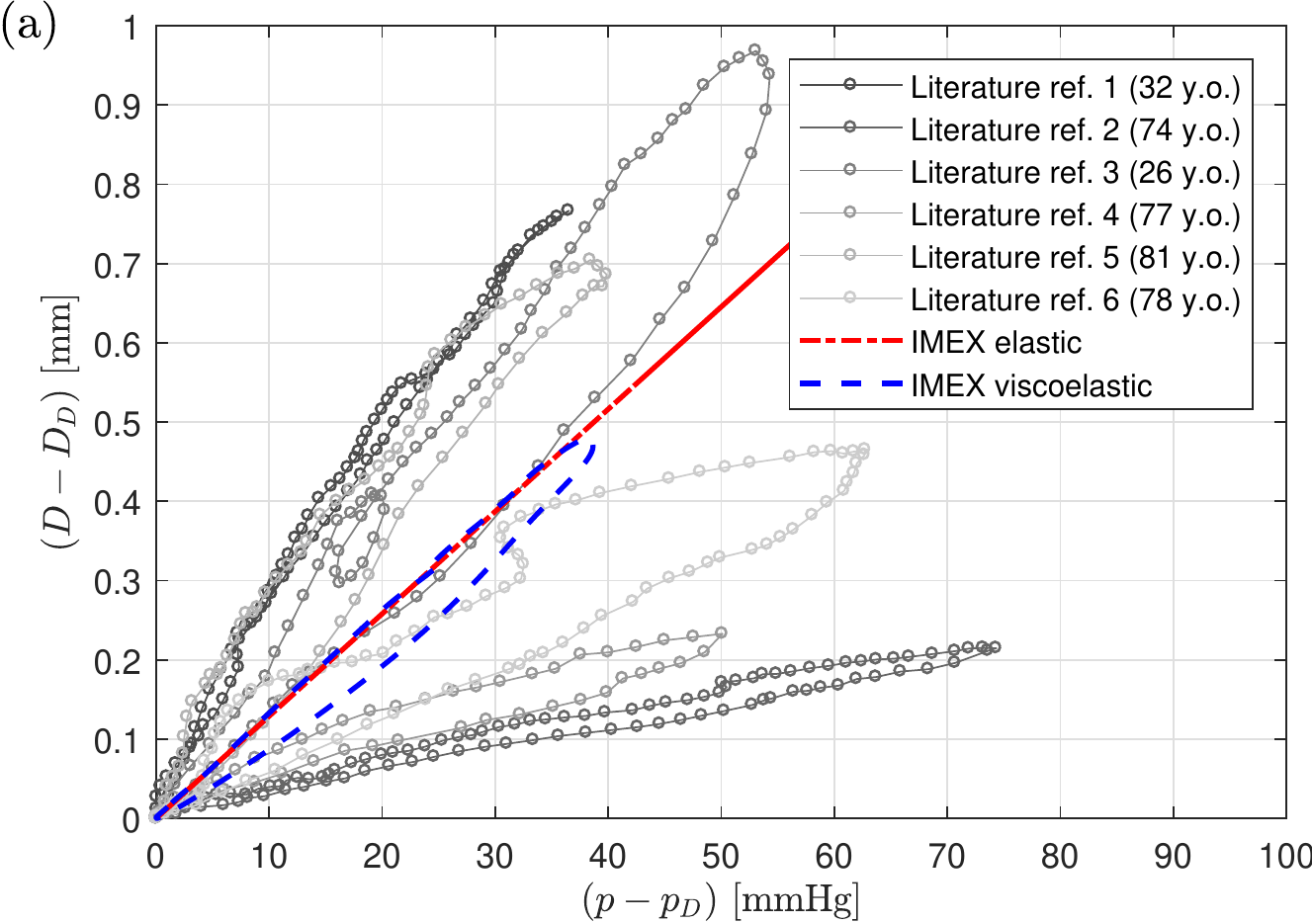}
\label{fig.CarotidNavas_hysteresis}
\vspace*{2mm}
\end{subfigure}
\begin{subfigure}{0.49\textwidth}
\centering
\includegraphics[width=0.95\linewidth]{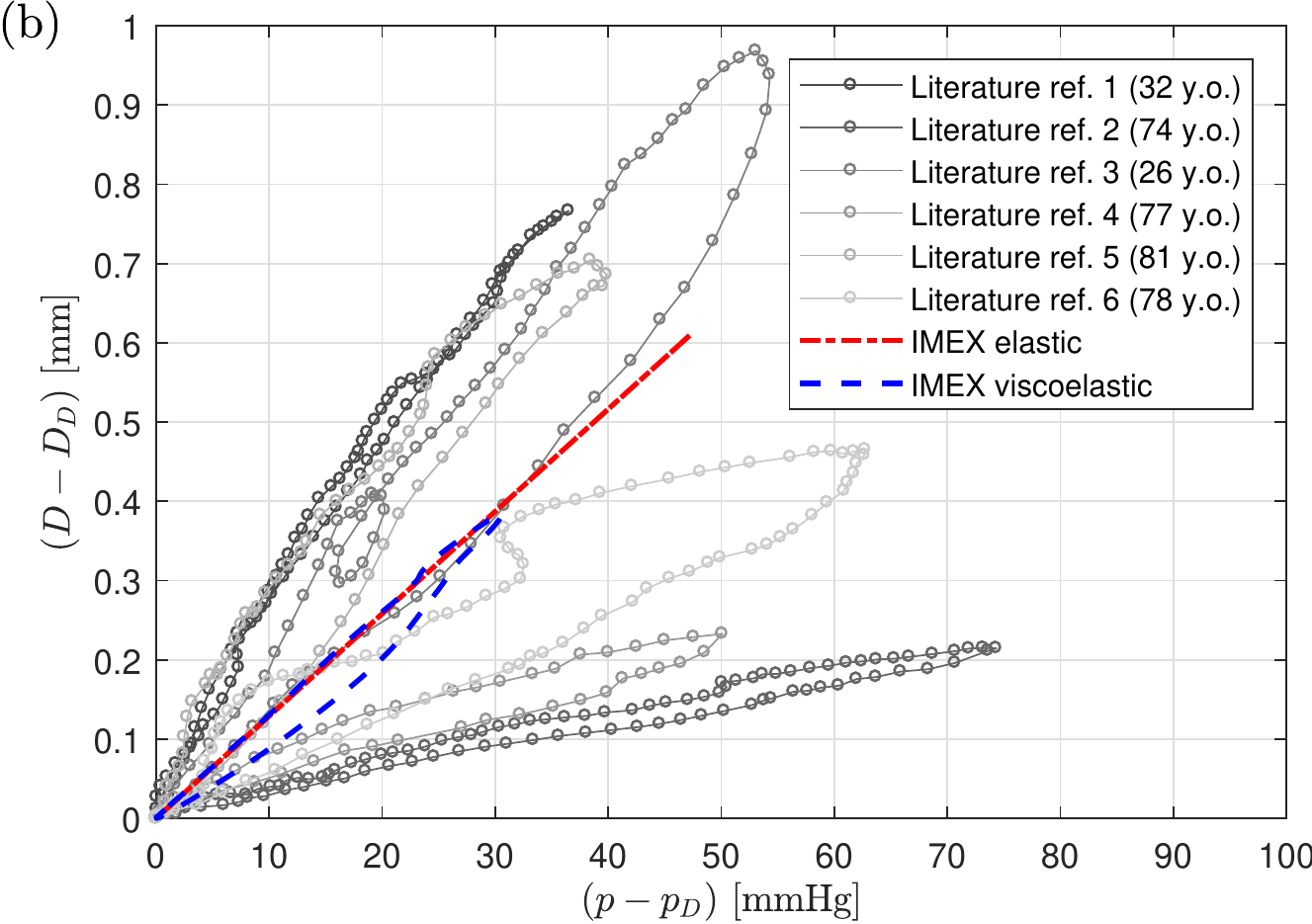}
\label{fig.CarotidMoreno_hysteresis}
\vspace*{2mm}
\end{subfigure}
\begin{subfigure}{0.49\textwidth}
\centering
\includegraphics[width=0.95\linewidth]{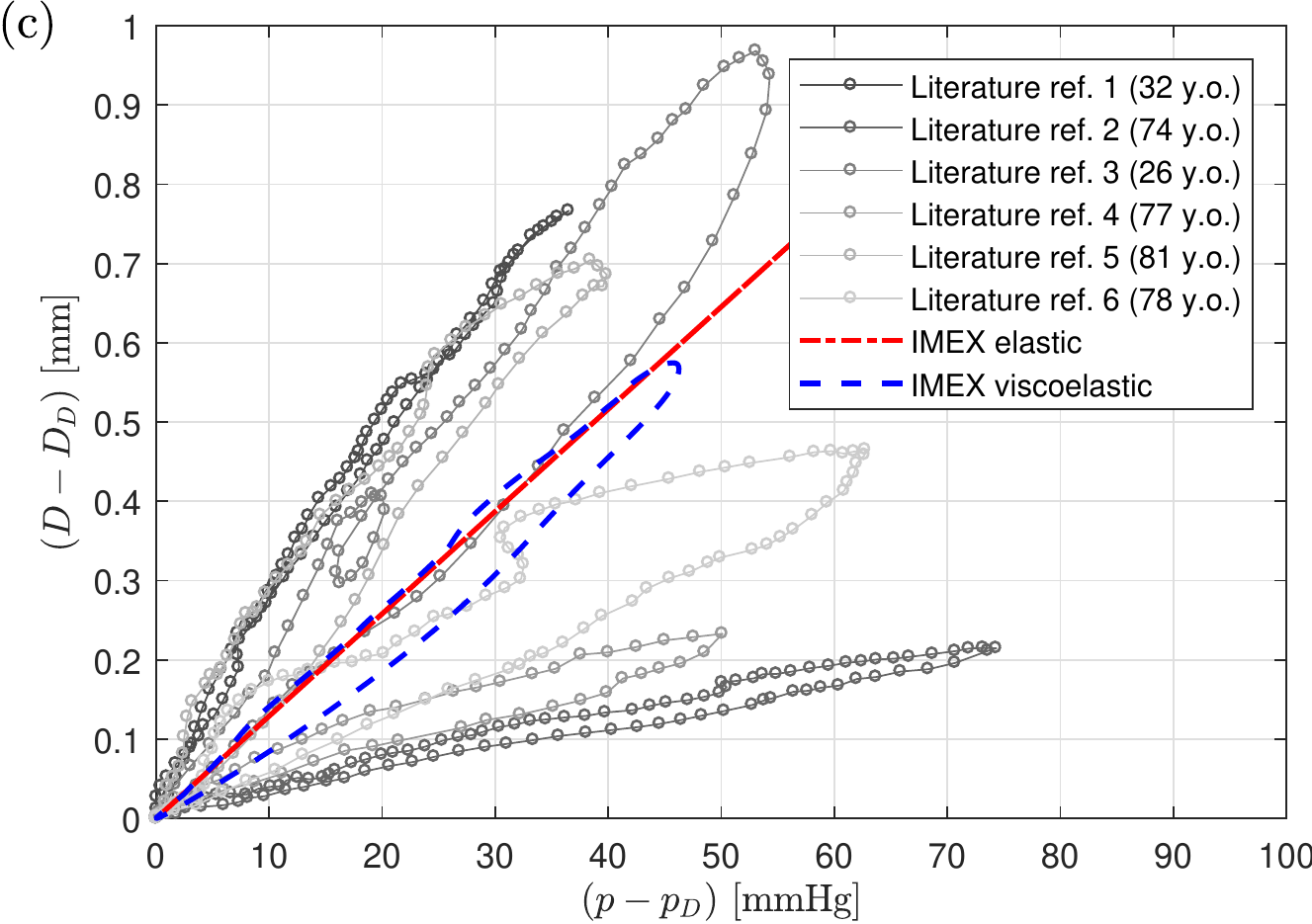}
\label{fig.CarotidRamos_hysteresis}
\vspace*{2mm}
\end{subfigure}
\begin{subfigure}{0.49\textwidth}
\centering
\includegraphics[width=0.95\linewidth]{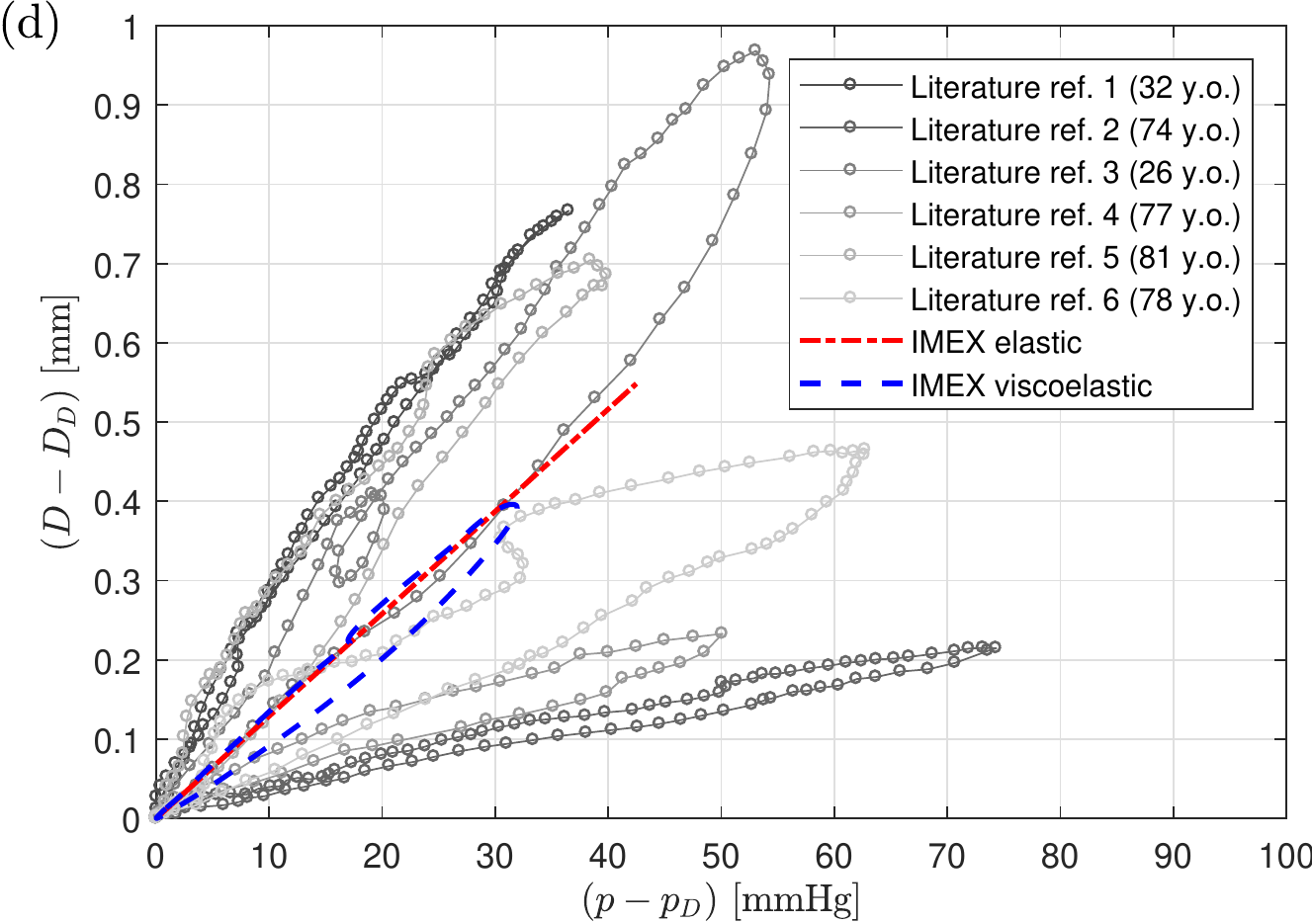}
\label{fig.CarotidBertaglia_hysteresis}
\vspace*{2mm}
\end{subfigure}
\begin{subfigure}{0.49\textwidth}
\centering
\includegraphics[width=0.95\linewidth]{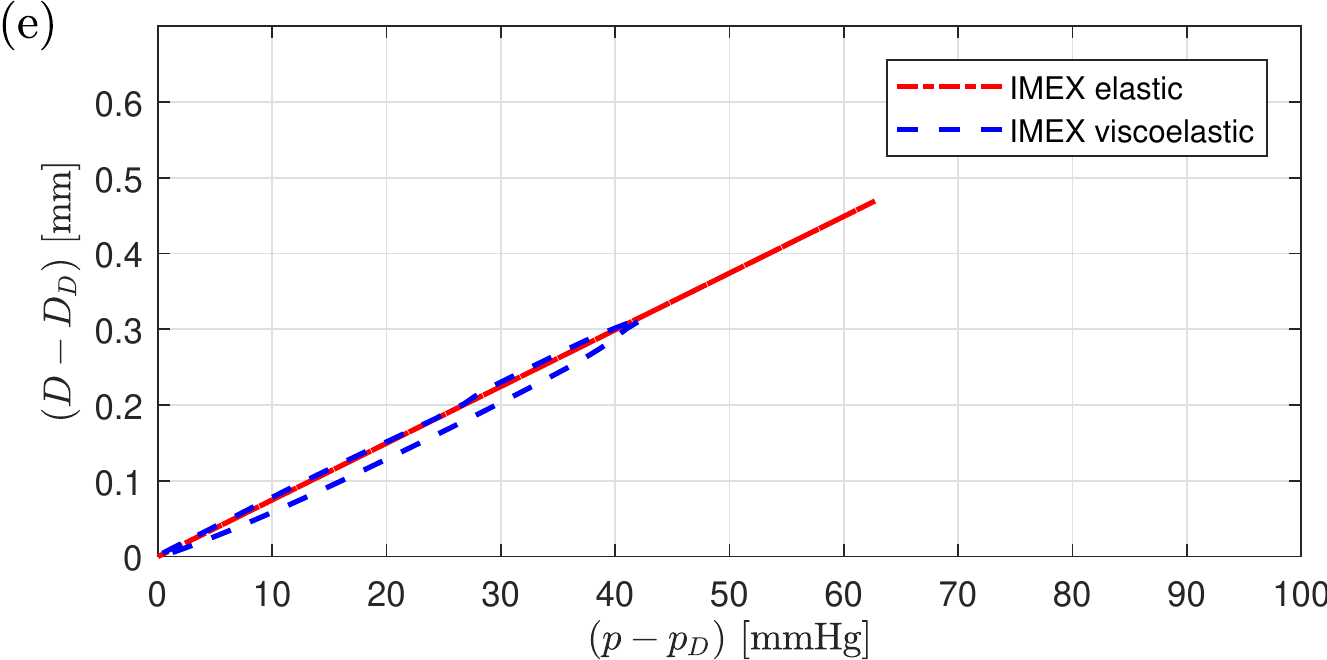}
\label{fig.FemoralSubj1_hysteresis}
\end{subfigure}
\hspace*{2mm}
\begin{subfigure}{0.49\textwidth}
\centering
\includegraphics[width=0.95\linewidth]{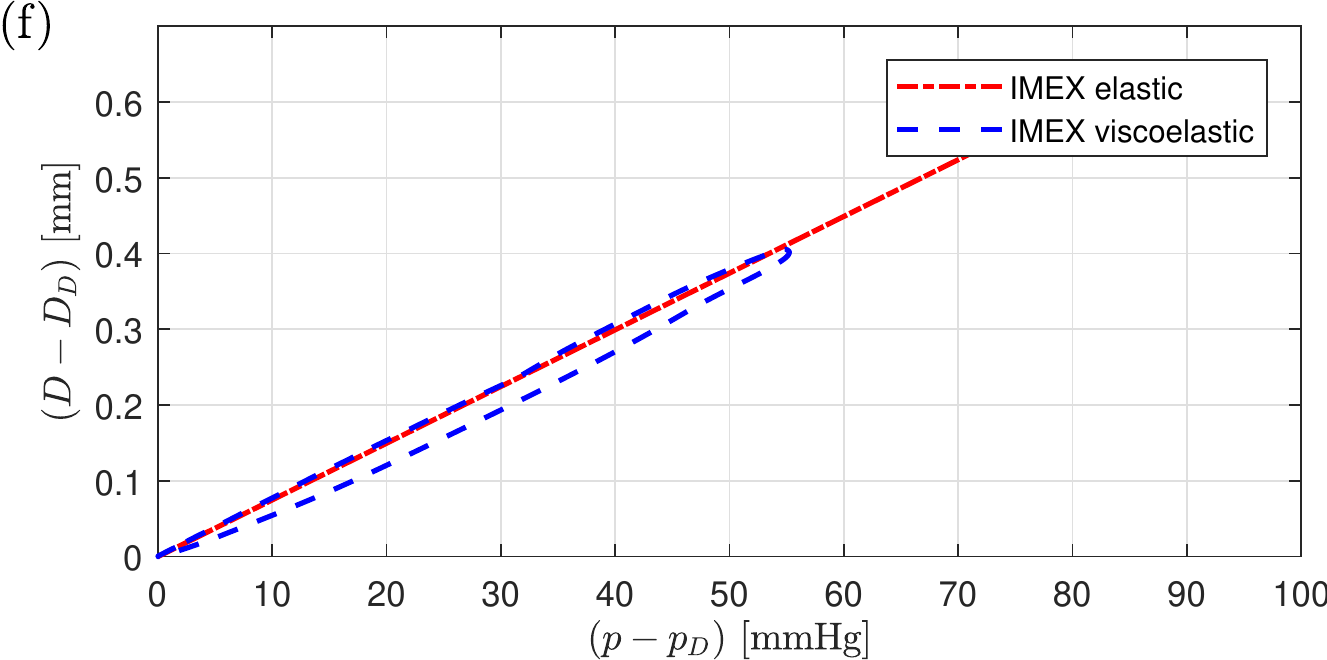}
\label{fig.FemoralSubj2_hysteresis}
\end{subfigure}
\caption{Hysteresis curves during one representative cardiac cycle presented in terms of relative pressure and relative diameter, with respect to diastolic values, $p_D$ and $D_D$ respectively. Results obtained solving the a-FSI system with the IMEX scheme with elastic and viscoelastic tube law for each subject of the study. Curves for the CCA of subjects A (a), B (b), C (c), D (d) and for the CFA of subjects E (e), F (f). Numerical results obtained for each carotid compared to hysteresis curves available in literature, produced correlating in-vivo diameter and pressure simultaneous records made with volunteers of different ages \cite{giannattasio2008,salvi2012}. Curves develop in time counter-clockwise. Numerical results obtained for the CFAs could not be compared to experimental data due to lack of sufficient data available in literature.}
\label{fig.hysteresis}
\end{figure}
\begin{table}[b!]
\centering
\begin{tabular}{l | c c c c c c }
\hline
	Parameter &CCA-A &CCA-B &CCA-C &CCA-D &FA-E &FA-F\\
	\hline
	Age	[years]	&29	&28	&44	&28	&44	&32\\
	\(L\) [cm] &17.70	&17.70	&17.70	&17.70	&14.50	&14.50 \\
	\(R_{0,in}\) [mm] &4.0	&4.0	&4.0	&4.0	&3.7	&3.7 \\
	\(R_{0,out}\) [mm] &3.7	&3.7	&3.7	&3.7	&3.14	&3.14 \\
	\(h_0\) [mm] &0.3	&0.3	&0.3	&0.3	&0.3	&0.3 \\
	\(p(x,0)\) [mmHg] &0	&0	&0	&0	&0	&0 \\
	\(u(x,0)\) [m/s] &0	&0	&0	&0	&0	&0 \\
	\(\alpha_c\) [-] &\sfrac{4}{3}	&\sfrac{4}{3}	&\sfrac{4}{3}	&\sfrac{4}{3}	&\sfrac{4}{3}	&\sfrac{4}{3} \\
	\(p_D\) [mmHg] &90.0	&70.0	&80.0	&75.0	&90.0	&90.0 \\
	\(p_{out}\) [mmHg] &0	&0	&0	&0	&0	&0 \\
	\(R_{1}\) [MPa s m$^{-3}$] &145.91	&145.91	&145.91	&145.91	&241.26 &241.26 \\
	\(R_{2}\) [MPa s m$^{-3}$] &768.17	&588.83	&702.28	&756.13	&4140.0	&2352.1\\
	\(C\) [m$^{3}$ GPa$^{-1}$] &0.29178	&0.24997	&0.49551	&0.11168	&0.11155	&0.13208 \\
	\(E_0\) [MPa] &1.7742	&1.7742	&1.7742	&1.7742 &2.2352 &2.2352 \\
	\(E_{\infty}\) [MPa] &0.9535	&0.9535	&0.9535	&0.9535	&1.2012	&1.2012 \\
	\(\eta\) [kPa s] &47.768	&47.768	&47.768	&47.768	&47.768	&47.768 \\
	\(\tau_r\) [s] &0.0125	&0.0125	&0.0125	&0.0125	&0.010	&0.010 \\
\hline
\end{tabular}
\caption{Model parameters of the common carotid artery (CCA) and common femoral artery (CFA) test cases for the 6 different subjects (A-F) from whom in-vivo velocity and pressure data were measured: subject age, vessel length $L$, inlet equilibrium radius $R_{0,in}$, outlet equilibrium radius $R_{0,in}$, vessel wall thickness $h_0$, initial pressure $p(x,0)$, initial velocity $u(x,0)$, Coriolis coefficient $\alpha_c$, diastolic pressure $p_D$ (in this model coincident with the external pressure $p_{ext}$ for the equilibrium), outflow pressure $p_{out}$, Windkessel resistance $R_1$, Windkessel resistance $R_2$, Windkessel compliance $C$, instantaneous Young modulus $E_0$, asymptotic Young modulus $E_{\infty}$, viscosity coefficient $\eta$, relaxation time $\tau_r$. Physiological data for arteries were taken from \cite{muller2014a}. Diastolic pressure was fixed based on values measured in the brachial artery for each patient. Outflow parameters were calibrated following the procedure proposed in \cite{alastruey2012a,xiao2014}. Viscoelastic parameters were estimated as presented in Section \ref{section_calibration}. In all the tests $\rho = 1060$~kg/m$^3$, $\mu = 0.004$~Pa~s. Concerning numerical parameters, $\CFL = 0.9$ in all the simulations, the number of cells in the domain is $nc = 7$ and  10 cardiac cycles are simulated, corresponding to a final time $t_{end} = 10.00$ s.}
\label{tab.datatestinvivo}
\end{table}
\subsection{In-vivo data test cases}
\label{section_TC_invivo}
Not having found benchmark test cases of blood flow in single viscoelastic vessels in literature, data measured in-vivo in human CCAs and CFAs (see Section \ref{section_invivodata}) are used as benchmark for the validation of the proposed model in its viscoelastic form. The velocity waveform extrapolated from each of the six subjects' Doppler measurements (four in the CCA and two in the CFA) is imposed at the inlet boundary ($u_{in}$) as described in \ref{section_BC}. The pressure waveform measured by the PulsePen tonometer is compared to the computed pressure obtained in the last cell of the domain, this being, for both type of vessels, the position that is considered closest to the data measurement position.  Figures \ref{fig.carotid_invivo} and \ref{fig.femoral_invivo} show flow rate and pressure results, obtained solving the a-FSI system considering the simple elastic and the SLS viscoelastic model with the IMEX scheme, in three different positions along the domain of each CCA and CFA respectively. It can be noticed that IMEX viscoelastic results correctly capture the shape and magnitude of the pressure waveform of all the volunteers concerning both the types of artery. These results confirm the capability of the proposed model to reproduce realistic pressure signals and the importance of taking into account the viscosity of the vessel wall in order not to overestimate systolic pressure values \cite{alastruey2011,battista2015,westerhof2004}.\\
In Fig.~\ref{fig.hysteresis}, computed $p$-$D$ hysteresis loops are presented, comparing those related to CCAs with reference literature hysteresis \cite{giannattasio2008,salvi2012}. It was not possible to do the same for CFA results, due to lack of sufficient reference data available in literature. Comparisons of the dissipated energy fractions, carried out for each subject from literature loops and computed hysteresis curves (as described in Section \ref{section_calibration}), confirm the correct reproduction of energy losses. Numerical dissipation percentage in CCA tests results 21.5\%, 20.0\%, 22.5\% and 23.1\%, respectively for subjects from A to D, while the average over age of the corresponding literature curves results 19.2\%, 23.7\%, 22.0\% and 23.7\%. Numerical results in CFA tests reproduce a dissipation percentage of 12.6\% and 13.7\%, respectively for subject E and F, asserting the presence of smaller areas of hysteresis in more peripheral vessels \cite{alastruey2012a}.
\section{Concluding remarks}
\label{section_conclusions}
In the present work, the a-FSI system for blood flow modeling is applied to real case studies in single arteries. Results obtained considering a simple elastic behavior of the vessel wall figure in perfect agreement with 1D and 3D benchmark data available in literature. In addition, the relevance of taking into account viscoelastic effects of arterial walls is confirmed comparing pressure results with experimental data collected from different human healthy volunteers in CCAs and CFAs. An effective procedure to estimate viscoelastic parameters of the SLS model is proposed, which returns CCA hysteresis curves dissipating energy fractions in line with values calculated from literature hysteresis loops in the same vessel. \\
Comparisons with in-vivo data demonstrate that the proposed model is able to correctly simulate pressure trends in different subjects, serving as a valuable tool to improve cardiovascular diagnostics and the treatment of diseases. Considering literature physiological data (for $R_0$, $h_0$, $L$ and $c_0$ in vessels), the procedure presented in Section \ref{section_calibration} permits to obtain all the necessary elastic and viscoelastic parameters. Only brachial systolic and diastolic pressure values (which can easily be recorded) are needed to correctly impose reference pressure values and define outlet lumped parameters \cite{alastruey2012a,xiao2014}, without the need of specific parameter adjustments.

\section*{Conflict of interest statement}
None.
\section*{Acknowledgements}
For this work, the authors Giulia Bertaglia, Valerio Caleffi and Alessandro Valiani were funded by MIUR (Ministero dell'Istruzione, dell'Universit\`a e della Ricerca) PRIN 2017 with the project \textit{``Innovative numerical methods for evolutionary partial differential equations and applications''}, code 2017KKJP4X. Valerio Caleffi was also funded by MIUR FFABR 2017. Adri\'an Navas-Montilla and Javier Murillo acknowledge the partial funding by Gobierno de Arag\'on through the Fondo Social Europeo.
\appendix
\renewcommand{\theequation}{\Alph{section}.\arabic{equation}}
\section{Boundary conditions}
\label{section_BC}
Referring to Fig.~\ref{fig.model}, inflow boundary conditions are defined prescribing the inlet flow rate $q_{in}$ or the inlet velocity $u_{in}$ (depending on the available data) and recurring to the first Riemann invariant associated with the genuinely non-linear fields:
\begin{equation}
\Gamma_1 = u - \int\frac{c}{A}\d A \hspace{0.5mm} .
\label{BC_RI1}
\end{equation}
The viscosity of vessels wall will be hereafter neglected when imposing boundary conditions, recurring to the simple elastic model relating $p$ to $A$ \cite{alastruey2012}.\\
At the outflow of the 1D domain, the RCR model, representative of the perfusion of downstream vessels, is coupled with the 1D model through the solution of the problem at the interface. A null outlet pressure, $p_{out} = 0$, which represents the pressure at which the flow arrives in the venous system \cite{xiao2014}, is prescribed. Discretizing system~\eqref{RCRsyst}, considering a time step size \(\Delta t = t^{k+1}-t^{k}\), leads to:
\begin{subequations}
\begin{align}
	&p_C^{k+1} - p_C^k = \frac{\Delta t}{C} \left[(q^*)^{k+1} - q_{out}^{k+1}\right] \label{BC_1.1} \\ 
	&(q^*)^{k+1} = \frac{1}{R_1} \left[p(A^*)^{k+1} - p_C^{k+1}\right] \label{BC_1.2} \\
	&q_{out}^{k+1} = \frac{1}{R_2} \left[p_C^{k+1} - p_{out}\right] . \label{BC_1.3}
\end{align}
\label{BC_1}
\end{subequations}
Subtracting Eq.~\eqref{BC_1.3} from Eq.~\eqref{BC_1.1} gives:
\begin{equation}
p_C^{k+1} = \frac{R_2}{\phi} (q^*)^{k+1} + \frac{\phi - 1}{\phi} p_C^{k} + \frac{p_{out}}{\phi} \hspace{0.5mm},
\label{pC}
\end{equation}
where $\phi = \frac{R_2C}{\Delta t} +1$. Using Eq.~\eqref{pC} into Eq.~\eqref{BC_1.2} the expression for $q^*$ is obtained, which reads:
\begin{equation}
(q^*)^{k+1} = \frac{p(A^*)^{k+1} - \tilde{p}}{R_1 + \frac{R_2}{\phi}} \hspace{0.5mm},
\label{q*}
\end{equation}
with $\tilde{p} = \phi^{-1}\left[p_{out} + (\phi - 1)(p_C)^k\right]$. Combining Eq.~\eqref{q*} with the second Riemann invariant associated with the genuinely non-linear fields,
\begin{equation}
\Gamma_2 = u + \int\frac{c}{A}\d A \hspace{0.5mm} ,
\label{BC_RI2}
\end{equation}
evaluated in the last cell and at the outlet interface of the 1D domain,
yields to a non-linear equation in $A^*$, solved using Newton's method \cite{alastruey2008}. Once $A^*$ is obtained, $u^*$ is calculated through $\Gamma_2$ and $p(A^*)$ is evaluated with the elastic tube law. \\
Note that when considering only the arterial network, the integral in the Riemann invariants \eqref{BC_RI1} and \eqref{BC_RI2} can be simplified, resulting:
\[\Gamma_1 = u - 4c, \qquad \Gamma_2 = u + 4c \hspace{0.5mm} .\]
\section{FSI parameters estimation algorithm}
\label{appendix_calib}

\begin{itemize}
\item Elastic model:
\begin{itemize}
\item Input: $\rho, R_0, h_0$ and $c_0$ from literature \cite{muller2014a,liang2009a,alastruey2012a,xiao2014},
\item $E_0$ evaluated with Eq.~\eqref{E_0},
\item $K$ evaluated with Eq.~\eqref{K}.
\end{itemize}
\item Viscoelastic model:
\begin{itemize}
\item Input: $\rho, R_0, h_0, c_0$ and $\Gamma$ from literature \cite{muller2014a,liang2009a,alastruey2012a,xiao2014},
\item $E_{\infty}$ evaluated adapting Eq.~\eqref{E_0} for the viscoelastic Young modulus: $E_{\infty} = \frac{2R_0 \rho c_0^2}{h_0}$,
\item $\eta$ evaluated with Eq.~\eqref{eta},
\item $E_0$ evaluated inverting Eq.~\eqref{Eratio_calib},
\item $K$ evaluated with Eq.~\eqref{K}.
\end{itemize}
\end{itemize}
\unappendix



\end{document}